\documentclass[conference]{IEEEtran}
\makeatletter
\def\ps@headings{%
\def\@oddhead{\mbox{}\scriptsize\rightmark \hfil \thepage}%
\def\@evenhead{\scriptsize\thepage \hfil \leftmark\mbox{}}%
\def\@oddfoot{}%
\def\@evenfoot{}}
\makeatother
\usepackage{cite}
\usepackage{graphicx}
\usepackage{amsmath, amsthm, amssymb}
\usepackage{algorithm,algorithmic}
\usepackage{enumerate}
\usepackage{balance, color}
\usepackage{multirow, verbatim}
\usepackage{url}
\usepackage{geometry}
\usepackage{tikz}
\usetikzlibrary{arrows}
\usetikzlibrary{shapes}
\usepackage{pgfplots}

\theoremstyle{plain} 
\newtheorem{definition}{Definition}
\newtheorem{theorem}{Theorem}
\theoremstyle{definition} 

\newcommand{\setup}{\ensuremath{\mathsf{HashSetup}}}
\newcommand{\hash}{\ensuremath{\mathsf{Hash}}}

\newcommand{\test}{\ensuremath{\mathsf{Test}}}
\newcommand{\vct}[1]{\ensuremath{\mathbf{#1}}}
\newcommand{\gnr}{\ensuremath{\mathsf{Generate}}}
\newcommand{\mac}{\ensuremath{\mathsf{Mac}}}
\newcommand{\cbn}{\ensuremath{\mathsf{Combine}}}
\newcommand{\vrf}{\ensuremath{\mathsf{Verify}}}
\newcommand{\sign}{\ensuremath{\mathsf{Sign}}}
\newcommand{\SMac}{\ensuremath{\mathsf{SpaceMac}}}
\newcommand{\IMac}{\ensuremath{\mathsf{InterMac}}}
\newcommand{\SHA}{\ensuremath{\mathsf{SHA\text{-}1}}}
\newcommand{\HMac}{\ensuremath{\mathsf{HMAC}}}
\renewcommand{\triangleq}{\ensuremath{\overset{\text{def}}{=}}}

\newcommand{\negl}{\ensuremath{\mathsf{negl}}}
\newcommand{\pp}{\ensuremath{\mathsf{pp}}}
\newcommand{\iden}{\ensuremath{\mathsf{id}}}

\newcommand{\HDL}{$\mathcal{H}$\text{-DL}}
\newcommand{\ie}{{\em i.e.}}
\newcommand{\ea}{{\em et al.}}
\newcommand{\eg}{{\em e.g.}}
\newcommand{\eff}[2]{\ensuremath{\mathbb{F}^{#1}_{#2}}}
\newcommand{\aug}[1]{\ensuremath{\mathsf{aug}(\vct{#1})}}

\newcommand{\lspan}[1]{\ensuremath{\mathsf{span}(\vct{#1})}}

\newcommand{\squishlist}{
 \begin{list}{$\bullet$}
  { \setlength{\itemsep}{0pt}
     \setlength{\parsep}{3pt}
     \setlength{\topsep}{3pt}
     \setlength{\partopsep}{0pt}
     \setlength{\leftmargin}{1.5em}
     \setlength{\labelwidth}{1em}
     \setlength{\labelsep}{0.5em} } }

\newcommand{\squishlisttwo}{
 \begin{list}{--}
  { \setlength{\itemsep}{0pt}
     \setlength{\parsep}{0pt}
    \setlength{\topsep}{0pt}
    \setlength{\partopsep}{0pt}
    \setlength{\leftmargin}{2em}
    \setlength{\labelwidth}{1.5em}
    \setlength{\labelsep}{0.5em} } }

\newcommand{\squishend}{
  \end{list}  }

\begin{document}

\title{On Detecting Pollution Attacks\\
in Inter-Session Network Coding}
\author{Anh Le, Athina Markopoulou\\University of California, Irvine\\\{\em {anh.le, athina\}@uci.edu}}
\date{}
\maketitle

%
%

\begin{abstract}
Dealing with pollution attacks in inter-session network coding is challenging due to the fact that sources, in addition to intermediate nodes, can be malicious. In this work, we precisely define corrupted packets in inter-session pollution based on the {\em commitment of the source packets}. We then propose three detection schemes: one hash-based and two MAC-based schemes: $\IMac_\text{CPK}$ and $\SMac_\text{PM}$. $\IMac_\text{CPK}$ is the first multi-source homomorphic MAC scheme that supports multiple keys. Both MAC schemes can replace traditional MACs, \eg, $\HMac$, in networks that employ inter-session coding. All three schemes provide in-network detection, are collusion-resistant, and have very low online bandwidth and computation overhead. 
\end{abstract}

\section{Introduction}
\label{sec:intro}
Network coding involves packets being combined at intermediate nodes inside the network. 
Depending on whether packets from the same or different sessions are mixed, network coding is classified as {\em intra-session} or {\em inter-session}, respectively. 
Inter-session coding, that is the focus of this paper, has been implemented in practice, such as in wireless mesh networks \cite{Katti2006, Omiwade2008} and streaming gestures \cite{Feng2011}.

The mixing nature of network coding makes it extremely vulnerable to {\em pollution} (a.k.a. Byzantine modification) attacks. In such an attack, malicious nodes inject corrupted packets that then are combined and forwarded by downstream nodes, eventually resulting in a large number of corrupted packets propagating in the network. This wastes network resources, such as bandwidth and CPU time. More critically, it prevents receivers from decoding the original packets. A large body of work has focused on pollution attacks in intra-session coding \cite{Cai2002, Zhang2006, Jaggi2007, Koetter2007, Ho2004, Kehdi2009, Yu2009, Dong2009, Gkantsidis2006, Li2006, Zhao2007, Charles2006, Boneh2009, Agrawal2009, Li2010, Jiang2010, Zhang2011, Jafarisiavoshani2008, Wang2010, Le2010, Le2011, Le2011a}, while pollution attacks in inter-session coding have received significantly less attention \cite{Agrawal2010, Yan2009, Dong2010}. 

In this paper, our goal is to detect pollution attacks in inter-session network coding using cryptographic primitives. 
This  is particularly challenging because not only intermediate nodes but also sources can be malicious and initiate attacks themselves. Recently, Agrawal \ea \cite{Agrawal2010} formulated the problem for the first time and presented a detection scheme based on homomorphic signatures. This scheme has high computation overhead due to many public-key signature verification and modular exponentiation operations performed at each node per packet. Furthermore, the signature size is large and does not scale as it increases linearly in the number of sources and packets sent by them.

In this paper, we introduce three novel detection schemes: one hash-based and two MAC- based schemes, all of which are significantly more efficient than \cite{Agrawal2010}. The key ingredient of our approaches is the use of {\em commitment (to a trusted controller) of source packets}. This commitment allows us to precisely define corrupted packets, thereby enabling detection of all corrupted packets, including some that \cite{Agrawal2010} cannot detect. We build upon this idea and design three schemes:
\squishlist
\item A hash-based detection scheme, that combines homomorphic \cite{Krohn2004} and traditional hash functions, \eg, $\SHA$.
\item  $\IMac_\text{CPK}$, a multi-source homomorphic MAC scheme. It is the first homomorphic MAC scheme that allows tags to be generated under different keys. 
\item $\SMac_\text{PM}$, a combination of an existing inner-product homomorphic MAC scheme (built for  intra-session coding \cite{Le2010}) and a private inner-product protocol \cite{Goethals2004}.
\squishend

Our hash-based scheme allows nodes to detect corrupted packets right after they receive them, thus providing {\em in-network detection}. Both of our MAC schemes can  replace traditional MACs, \eg, $\HMac$, to provide {\em end-to-end detection}. Moreover, they can be used as building blocks for other schemes that provide in-network detection, such as \cite{Agrawal2009, Li2010, Zhang2011} and \cite{Le2011a}. The hash-based detection scheme is arbitrarily {\em collusion-resistant}. Meanwhile, depending on the in-network detection scheme used, a scheme built on one of the MAC schemes could be either arbitrarily collusion-resistant or $c$-collusion-resistant, for a predetermined small $c$. We also custom design commitment schemes that offer high bandwidth efficiency for both MAC schemes. Most importantly, all proposed schemes have significantly higher {\em bandwidth} and {\em computation efficiency} than those of the state-of-the-art detection scheme for inter-session coding \cite{Agrawal2010}. In particular, simulation results show that for a detection scheme built on one of our MAC schemes, both the online bandwidth and computation overhead are low, as low as 3\% and 4 ms, respectively.


The proposed schemes provide alternative approaches to detect corrupted packets in inter-session network coding. In general, the MAC-based schemes have significantly lower computation overhead than the hash-based scheme (Section \ref{subsec:computationOverhead}). $\SMac_\text{PM}$ offers lower commitment overhead (Section \ref{subsec:bandwidthOverhead}), but $\IMac_\text{CPK}$ is less vulnerable to colluding malicious receivers (end of Section \ref{subsec:privateMAC}).

The rest of this paper is organized as follows. In Section \ref{sec:related_work}, we discuss related work. In Section \ref{sec:formulation}, we describe the network operations, threat models, and definition of corrupted packets. 
In Section \ref{sec:hash_based}, we present the proposed hash-based detection scheme. In Section \ref{sec:mac_based}, we describe $\IMac_\text{CPK}$ and $\SMac_\text{PM}$. In Section \ref{sec:evaluation}, we evaluate the performance of our schemes. Finally, we conclude in Section \ref{sec:conclusion}.

\section{Related Work}
\label{sec:related_work}

Because pollution attacks pose a severe threat to the success of network coding, a large body of research has been devoted to designing defense mechanisms, including  both information theoretic and cryptographic approaches. The existing approaches provide error-correction capability \cite{Cai2002, Zhang2006, Jaggi2007, Koetter2007}, attack detection \cite{Ho2004, Kehdi2009, Yu2009, Dong2009, Gkantsidis2006, Li2006, Zhao2007, Charles2006, Boneh2009, Agrawal2009, Li2010, Zhang2011, Le2011, Le2011a, Agrawal2010}, and attacker identification \cite{Jafarisiavoshani2008, Wang2010, Le2010, Dong2010}. Most of these approaches, including our prior work \cite{Le2010, Le2011, Le2011a}, are proposed for intra-session coding and are not applicable to inter-session coding, as discussed in Section \ref{subsec:threat_models}. We refer the reader to \cite{Le2011} for a comprehensive overview of intra-session defense mechanisms. Here, we focus on defense against pollution attacks in inter-session network coding.


Agrawal \ea \cite{Agrawal2010} proposed a homomorphic signature scheme to provide in-network detection for inter-session network coding. In their scheme, the signature of a packet sent by a source $S$ consists of $g$ hash values of all $g$ source packets sent by $S$, together with the public key signature of the hash values. The hash values are computed using a homomorphic hash function proposed in \cite{Krohn2004}. The signature of the hash is computed using a secure signature scheme. 
The signature $\sigma_\vct{y}$ of a packet $\vct{y}$, which is a linear combination of packets belonging to $\ell$ different flows, is the concatenation of the signatures of $\ell$ different signatures. 
The main drawbacks of this scheme are (i) the expensive verification: the verification of $\sigma_\vct{y}$ involves $\ell$ public-key signature verification and one homomorphic hash verification, and (ii) the large signature size: the size of $\sigma_\vct{y}$ is large, including $\ell$ public-key signatures and $g \ell$ hash values.

The approaches proposed in this paper are inherently different from \cite{Agrawal2010}. We leverage the commitment of source packets and build our detection schemes based on un-key and symmetric-key cryptographic primitives as opposed to public-key primitives. We significantly improve the bandwidth and computation efficiency over \cite{Agrawal2010} (Section \ref{sec:evaluation}). Furthermore, by precisely defining corrupted packets, our schemes are able to detect some corrupted packets that \cite{Agrawal2010} cannot (Section \ref{subsec:corrupted}).


Dong \ea \cite{Dong2010} proposed a scheme that allows for identifying malicious nodes in inter-session network coding. 
When a pollution is detected, a bit-level traceback procedure is executed to identify the attacker. 
Our detection schemes are orthogonal and complementary to this identification scheme.

\section{Problem Formulation}
\label{sec:formulation}

\subsection{Network Model and Operation}
\label{subsec:network_model}
Some of the notation we use are from \cite{Le2011} and \cite{Agrawal2010}. Consider a graph denoted by $\mathcal{G} = (\mathcal{V}, \mathcal{E})$. There are $s$ pairs of source-receiver in the network, denoted by $(S_i, R_i), i \in [1, s]$. Each source, $S_i$, sends packets to its corresponding receiver, $R_i$, by first dividing the packets into generations. For simplicity, we assume that all sources use the same generation size, $g$. It is straightforward to extend our defense schemes to accommodate different generation sizes. $S_i$ interprets its packets in a single generation, $\hat{\vct{v}}_{ij}, j \in [1,g]$, as vectors in a $n$-dimensional vector space over a finite field $\eff{}{q}$. Before sending, $S_i$ appends to $\hat{\vct{v}}_{i,j}$ its coding coefficient, forming $g$ {\em augmented packets}, $\vct{v}_{i,1}, \cdots, \vct{v}_{i,g}$:
{\addtolength{\belowdisplayskip}{-7mm} 
\addtolength{\abovedisplayskip}{-6mm} 
\begin{align*}
~&\vct{v}_{i,j} = (\textrm{---}\vct{\hat{v}}_{i,j}\textrm{---}, \underbrace{0, \cdots, 0}_{g \times (i-1)}, \underbrace{\overbrace{0, \cdots, 0, 1}^j, 0, \cdots, 0}_g, \underbrace{0, \cdots, 0}_{g \times (s-i)})\,.
\end{align*}}
{\flushleft We} refer to the augmented packets, $\vct{v}_{i,j}$'s, as {\em source packets} and $\hat{\vct{v}}_{i,j}$ as data of $\vct{v}_{i,j}$. We use $\mathsf{aug}(\vct{v}_{i,j})$ to denote the coding coefficients of $\vct{v}_{i,j}$.

Note that for each generation, there are $m \triangleq sg$ source packets. The sources send source packets into the network generation by generation. Intermediate nodes in the network perform generation-based linear network coding, \ie, they linearly combine packets that belong to the same generation. Packets sent from different sources may be combined by intermediate nodes. For example, when an intermediate node $N$ receives $\ell$ packets, $\vct{w}_1, \cdots, \vct{w}_\ell$, which are some linear combinations of the source packets sent by any set of sources, it chooses $\ell$ {\em local coding coefficients}, $\alpha_1, \cdots, \alpha_\ell$, depending on the coding scheme used, and then transmit $\vct{y} = \sum_{i=1}^\ell \alpha_i \vct{w}_i$ to one or more of its outgoing edges. Note that if $\vct{y}$ is a linear combination of the source packets $\vct{v}_{i,j}$'s then the last $m$ symbols of $\vct{y}$ contain its {\em global coding coefficients}. For clarity, we focus on the transmission of a single generation by all the sources.

Let the subspace spanned by the source packets be $\Pi \triangleq \mathsf{span}(\vct{v}_{1,1}, \cdots, \vct{v}_{s,g})$ and the subspace spanned by the data of the source packets be $\hat{\Pi} \triangleq \mathsf{span}(\hat{\vct{v}}_{1,1}, \cdots, \hat{\vct{v}}_{s,g})$. We refer to $\Pi$ as the {\em source space} and $\hat{\Pi}$ as the {\em source data space}. When all nodes in the networks are benign, all packets in the network belong to the source space. A receiver, $R_i$, can decode the original packets sent by its corresponding source $S_i$ after collecting enough packets. In particular, after collecting $m$ linearly independent packets, $R_i$ can decode the original packets by applying Gaussian elimination on the $m \times (n+m)$ matrix formed by the collected packets. $R_i$ may also be able to decode using less than $m$ linearly independent packets because $R_i$ is not interested in packets sent by the other sources. 

\subsection{Inter-Session Network Coding Characteristics}
\label{subsec:interSession}

In inter-session network coding, it is often the case that intermediate nodes are able to decode source packets from the received coded packets. For instance, in COPE \cite{Katti2006}, every encoded packet is decoded at the next hop. There are also other coding schemes where encoded packet are decoded by either the first hop or the second hop, \eg, see  \cite{Omiwade2008} and \cite{Feng2011}. Furthermore, in inter-session coding, source packets of a source $S_i$ may not traverse  the whole network but only some parts of the network: for instance, in a directed acyclic graph, packets sent from $S_i$ should not travel to nodes that have no path to $R_i$. We will exploit these observations later in the proposed schemes.

Finally, the most important observation is that, in inter-session network coding, not only intermediate nodes but also some sources may be malicious. This differentiates the scenario we study in this work from single-source intra-session coding. We explicitly take this observation into account in our threat model below.

\subsection{Threat Model}
\label{subsec:threat_models}

We assume that up to $s-1$ sources could be malicious, any intermediate node may be malicious,  and the receivers are trusted. To pollute the network, the malicious nodes may generate and inject any type of traffic into the network; they may also collude among themselves. We assume the attackers know about the construction of any cryptographic primitive used but the attackers' running time is polynomial in the security parameter of cryptographic primitives.

\begin{figure}[tp]
\centering
\begin{tikzpicture}
\node[scale=0.8] {
\begin{tikzpicture}
\tikzstyle{circ} =[circle,draw=black!50,fill=black!10,thick]
\tikzstyle{circX} =[circle,draw=black!50,fill=red!60,thick]
\tikzstyle{pre} =[<-,shorten <=1pt,>=stealth',semithick]
\tikzstyle{post}=[->,shorten >=1pt,>=stealth',semithick]
	\node[circ]		(nS1)		at (-4, 1)		{\footnotesize $S_1$};
	\node[circX]		(nS2)		at ( 4, 1)		{\footnotesize $S_2$};
	\node[circ] 		(nA)			at ( 0, 1)		{$A$}
		edge [pre] node[midway,sloped,above] 	{$(\hat{\vct{v}}_1,1,0)$} (nS1)
		edge [pre] node[midway,sloped,above] 	{$(\hat{\vct{v}}'_2,0,1)$} (nS2);
	\node[circ] 		(nB)			at ( 0,-1)		{$B$}
		edge [pre] node[auto,swap] 			{$(\hat{\vct{v}}_1+\hat{\vct{v}}'_2,1,1)$} (nA);
	\node[circ] 		(nR2)		at (-4,-1)		{\footnotesize $R_2$}
		edge [pre] node[auto,swap] 			{$(\hat{\vct{v}}_1,1,0)$} (nS1)
		edge [pre] node[midway,sloped,above] 	{$(\hat{\vct{v}}_1+\hat{\vct{v}}'_2,1,1)$} (nB);
	\node[circ] 		(nR1)		at ( 4,-1)		{\footnotesize $R_1$}
		edge [pre] node[auto,swap] 			{$(\hat{\vct{v}}_2,0,1)$} (nS2)
		edge [pre] node[midway,sloped,above] 	{$(\hat{\vct{v}}_1+\hat{\vct{v}}'_2,1,1)$} (nB);
\end{tikzpicture}
};
\end{tikzpicture}
\vspace*{-3mm}
\caption{An example of pollution attack in inter-session network coding. Source $S_2$ is malicious and all other nodes are benign. $S_2$ pollutes the flow $S_1$-$R_1$ by injecting conflicting source packets $(\hat{\vct{v}}'_2,0,1)$ and $(\hat{\vct{v}}_2,0,1)$. $R_1$ decodes and recovers incorrect $\hat{\vct{v}}_1$.}
\label{fig:multi-src-atk}
\vspace*{-6mm}
\end{figure}
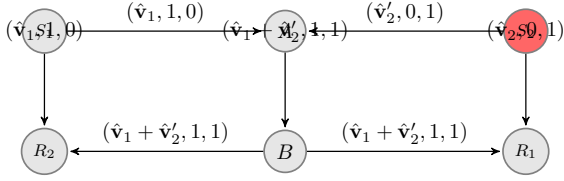

{\flushleft \bf Example Attack.}\quad Fig. \ref{fig:multi-src-atk} depicts the classic butterfly network coding across two unicast sessions. There are two sources: $S_1$ is benign, but $S_2$ is malicious. $A$, $B$, $R_1$, and $R_2$ are benign. The generation size is 1. Only node $A$ combines incoming packets, and only $R_1$ and $R_2$ decode. Local coding coefficients at $A$ are fixed: $\alpha_1 = \alpha_2=1$. Packets sent by the nodes are annotated on the edges. In this example, $S_2$ successfully pollutes the network because it causes an incorrect decoding at $R_1$. More specifically, by subtracting $(\hat{\vct{v}}_2,0,1)$ from $(\hat{\vct{v}}_1+\hat{\vct{v}}'_2,1,1)$, $R_1$ receives $\hat{\vct{v}}_1+\hat{\vct{v}}'_2-\hat{\vct{v}}_2$ instead of $\hat{\vct{v}}_1$.

{\flushleft \bf Intra-Session Detection Failure.}\quad 
Both unkey cryptographic approaches and key-based cryptographic approaches developed for intra-session fail to detect corrupted packets in the inter-session threat model. The ways they fail, however, are different. We first consider applying the hash-based scheme proposed in \cite{Gkantsidis2006}. Prior to the transmission, $A$, $B$, $R_1$, and $R_2$ download the hash of $\hat{\vct{v}}_1$ from $S_1$ and hash of $\hat{\vct{v}}_2$ from $S_2$. $S_2$ can act maliciously by sending to $R_1$ the hash of $\hat{\vct{v}}_2$ but sending to $A$ and $B$ the hash of $\hat{\vct{v}}'_2$. This makes $A$ accept $(\hat{\vct{v}}'_2,1,0)$, $B$ accept $(\hat{\vct{v}}_1+\hat{\vct{v}}'_2,1,1)$, and $R_1$ accept $(\hat{\vct{v}}_2,1,0)$. Therefore, $S_2$ can still carry out the same attack.

Now, let us consider applying any of the proposed MAC or signature-based approaches, such as, \cite{Charles2006, Boneh2009, Yu2009, Agrawal2009, Li2010, Zhang2011, Le2011}. When using any one of these schemes, MAC tags or signatures of packets must be generated under the same (private or symmetric) secret key so that the homomorphic property of the scheme holds. But if this is the case, a malicious source knowing the key can generate a valid tag/signature of any packet of its interest and pollute the network. For instance, $S_2$ can send to $R_1$ $(\hat{\vct{v}}'_1,1,0)$ and its valid tag/signature, where $\hat{\vct{v}}_1 \neq \hat{\vct{v}}'_1$, and $R_1$ will accept this corrupted packet.


\subsection{ Corrupted Packet}
\label{subsec:corrupted}
Loosely speaking, we consider any packet that causes a pollution of flows from benign sources corrupted. Nevertheless, in order to detect a pollution attack, corrupted packets must be precisely defined. We first require that each source, $S_i$, {\em commits} to its source packets before the transmission. We then define a corrupted packet based on this commitment. (i) In our hash-based scheme, we require each source to commit to the data of each of its packets by sending the hash of the data to a trusted controller. Let $\hat{\Pi}$ be the space spanned by the committed data of all the sources. We call $\hat{\Pi}$ the {\em committed source data space}. (ii) In our MAC-based schemes, we require each source to commit to each of its whole packet as opposed to just the data. We call the space spanned by all the committed source packets the {\em committed source space} and denote it by $\Pi$.

\begin{definition}
Let $\hat{\Pi}$ and $\Pi$ be the committed source data space and committed source space, respectively. A packet $\vct{y}$ is considered {\em corrupted} if $\hat{\vct{y}} \notin \hat{\Pi}$ or $\vct{y} \notin \Pi$.
\end{definition}

The above definition helps us to design detection schemes capable of detecting all corrupted packets. For instance, in Fig. \ref{fig:multi-src-atk}, if $S_2$ commits to $\hat{\vct{v}}_2$ then our schemes will help nodes $A$ to drop $(\hat{\vct{v}}'_2,0,1)$, thus avoiding having $(\hat{\vct{v}}_1+\hat{\vct{v}}'_2,1,1)$. In contrast, the scheme in \cite{Agrawal2010} only helps a node to detect conflicting packets and does not detect all corrupted packets. For instance, if \cite{Agrawal2010} is used, $A$ and $B$ still accept $\hat{\vct{v}}'_2$ and $(\hat{\vct{v}}_1+\hat{\vct{v}}'_2,1,1)$, respectively. $(\hat{\vct{v}}_1+\hat{\vct{v}}'_2,1,1)$ is detected as corrupted at $R_1$ if $R_1$ receives $\hat{\vct{v}}_2$ first, or $\hat{\vct{v}}_2$ is detected as corrupted if $R_1$ receives $(\hat{\vct{v}}_1+\hat{\vct{v}}'_2,1,1)$ first.

\subsection{Trusted Controller}
Trusted controllers have been used explicitly in previous work that identify and eliminate attackers \cite{Jafarisiavoshani2008, Wang2010, Le2010}. They have also been introduced implicitly by other detection schemes \cite{Kehdi2009, Yu2009, Dong2009, Gkantsidis2006, Li2006, Zhao2007, Charles2006, Boneh2009, Agrawal2009, Li2010, Jiang2010, Zhang2011}, where a {\em trusted source} setups and distributes hash values, MAC tags, and keys. In this work, we explicitly uses a standalone trusted controller to support the commitment.

\section{The Hash-Based Detection}
\label{sec:hash_based}

\subsection{Key Observations and Approach}
{\flushleft \bf Observation 1.}\quad Let us revisit the discussion of applying homomorphic hash functions to inter-session network coding in Section \ref{subsec:threat_models}. We observe that the main reason why $S_2$ can successfully pollute flow $S_1$-$R_1$ is that $S_2$ is able to distribute different hash values of $\hat{\vct{v}}_2$ and $\hat{\vct{v}}'_2$ to $A$, $B$, and $R_1$. If all nodes in the network receive the same hash value, either hash of $\hat{\vct{v}}_2$ or $\hat{\vct{v}}'_2$, then $S_2$ will not be able to carry out the attack because one of the two will be dropped due to incorrect hash. Ensuring that all nodes in the network receive the same hash value of $\hat{\vct{v}}_2$ or $\hat{\vct{v}}_2'$ is in fact equivalent to forcing $S_2$ to commit to either $\hat{\vct{v}}_2$ or $\hat{\vct{v}}'_2$, thus making any linear combination involving the other (non-committed) packet a corrupted packet. 

{\flushleft \bf Observation 2.}\quad As mentioned in Section \ref{subsec:interSession}, in inter-session network coding, it is often the case that intermediate nodes completely decode coded packets and recover their corresponding source packets. We exploit this fact and propose to use traditional hash functions to check for the integrity of these decodable packets. In other words, instead of verifying a coded packet using an expensive homomorphic hash verification, a node decodes it and verifies the recovered packet using an inexpensive traditional hash verification. Note that a traditional hash verification is two to three orders of magnitude less expensive than a homomorphic one. This observation is especially beneficial to COPE-like coding schemes \cite{Katti2006}, where every coded packet is decodable by any next hop.

\begin{figure}[t]
\centering
\begin{tikzpicture}
\tikzstyle{circ} =[circle,draw=black!50,fill=black!10,thick]
\tikzstyle{circX} =[circle,draw=black!50,fill=green!40,thick]
\tikzstyle{pre} =[<-,shorten <=2pt, >=stealth',semithick,dashed,draw=black]
\tikzstyle{post}=[->,shorten >=2pt, >=stealth',semithick,dotted,draw=blue]
	\node[circX]	(C)			at (3,0)			{$C$};
	\node[circ]	(S1)		at (0, 0)			{\footnotesize $S_1$}
		edge[post] node[blue,above] {$(h_{1,1}, \bar{h}_{1,1})$} (C) ;
	\node[circ]	(S2)		at (6, 0)			{\footnotesize $S_2$}
		edge[post] node[blue,above] {$(h_{2,1}, \bar{h}_{2,1})$} (C) ;
	\node[circ]	(A)			at (0, -2)		{$A$}
		edge[pre] node[midway,sloped,above] {$h_{i,j}, \bar{h}_{i,j}$} (C) ;
	\node[circ]	(B)			at (2, -2)		{$B$}
		edge[pre] (C) ;
	\node[circ]	(R1)		at (4, -2)		{\footnotesize $R_1$}
		edge[pre] (C) ;
	\node[circ]	(R2)		at (6, -2)		{\footnotesize $R_2$}
		edge[pre] node[midway,sloped,above] {$h_{i,j}, \bar{h}_{i,j}$} (C) ;
\end{tikzpicture}
\caption{Commitment and hash distribution for the network of Fig. \ref{fig:multi-src-atk}.}
\label{fig:hashNetwork}
\end{figure}
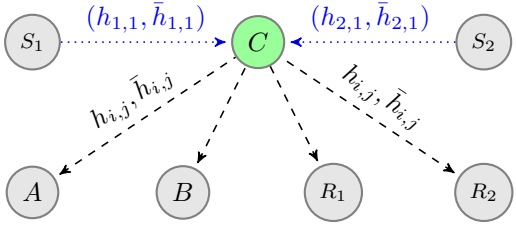

{\flushleft \bf Approach.}\quad Our hash-based detection scheme needs a trusted controller. Denote this controller by $C$. The scheme is based on the above observations and works as follows:

{\flushleft \em Setup:} $C$ sends to every node the description of a homomorphic hash function (\eg, \HDL, described in the next section) as well as a traditional hash function, \eg, $\SHA$. Before sending, each source, $S_i~(i \in [1,s])$, augments its data following the augmentation scheme described in section \ref{sec:formulation}. For every source packet, $\vct{v}_{ij}~(j \in [1,g])$, $S_i$ computes a homomorphic hash value and a traditional hash value, denoted as $h_{ij}$ and $\bar{h}_{ij}$, respectively. Each source then sends both $h_{ij}$ and $\bar{h}_{ij}$ to $C$. The commitment of each source are the pairs $(h_{ij}$, $\bar{h}_{ij})$. Every node downloads these pairs from $C$. We assume that the hash descriptions and values are distributed through authentic (tampering resistant) channels as usual applications of hash. Fig. \ref{fig:hashNetwork} illustrates how the hashes are distributed for the network of Fig. \ref{fig:multi-src-atk}.

{\flushleft \em Sending:} At each node, sending packets, including linearly combining incoming packets, is performed as usually.
{\flushleft \em Receiving and Verification:} Upon receiving a packet $\vct{y}$, if a node is specified to decode by the coding scheme, it checks if it can recover a source packet by decoding using $\vct{y}$ and its previously received packets. (i) If it can, it uses the traditional hash check to verify the integrity of the packet. (ii) If it cannot or in the case the node is not specified to decode, it uses the homomorphic hash check to verify the integrity of $\vct{y}$. If the recovered source packet (case (i)) or $\vct{y}$ (case (ii)) passes the verification, the node marks $\vct{y}$ as legitimate and uses it in subsequent transmissions; otherwise, it drops $\vct{y}$.

\subsection{Homomorphic Hash Scheme}
A homomorphic hash scheme consists of three polynomial-time algorithms:
\squishlist
\item $\setup (1^\lambda,n)$: Input: unary representation of the security parameter $\lambda$, and the dimension of the data space $n$. Output: public parameters $\pp$.

\item $\hash (\pp, \hat{\vct{v}}):$ Input: public parameters $\pp$ and a data vector $\hat{\vct{v}} \in \eff{n}{q}$. Output: hash value, $h \in \eff{}{q}$, of $\hat{\vct{v}}$.\\
-- The hash of $\hat{\vct{y}}$, a linear combination of $m$ source data vectors $\hat{\vct{v}}_i, i \in [1,m]$, is a hash vector $\vct{h} = (h_1, \cdots, h_m)$, where $h_i = \hash (\pp, \hat{\vct{v}}_i)$.

\item $\test (\pp, \hat{\vct{y}}, \bar{\beta}, \vct{h}):$ Input: public parameters $\pp$, a vector  $\hat{\vct{y}} \in \eff{n}{q}$, a vector of coefficient $\bar{\beta} \in \eff{m}{q}$, and a hash vector $\vct{h} \in \eff{m}{q}$. Output: $\top$ (true) or $\bot$ (false).
\squishend
Intuitively, let $\vct{h}$ be the set of hashes of the data of the source packets. For a packet $\vct{y}$ with data $\hat{\vct{y}}$ and coding coefficients $\bar{\beta}$, if $\vct{y}$ is a linear combination of the source packets then $\test$ should outputs $\top$. Also, it should be difficult for an adversary to find a packet $\vct{y}$ outside of the source space such that $\test$ outputs $\top$.

{\flushleft \bf Correctness.}\quad For all $\pp \leftarrow \setup (1^\lambda,n)$, we require the following properties for the correctness of the scheme:
\squishlist
\item For all $\hat{\vct{v}} \in \eff{n}{q}$, if $h = \hash(\pp, \hat{\vct{v}})$ then for all $i \in [1,m]$, $\test(\pp, \hat{\vct{v}}, \vct{e}_i, \vct{h}) = \top\,,$ where $\vct{e}_i$ is the $i$-th unit vector of the space $\eff{m}{q}$ and the $j$-th component of $\vct{h}$, $\vct{h}^{(j)}$, is defined as follows: $\vct{h}^{(j)}$ equals $h$ if $j$ equals $i$ and equals $r_j$ otherwise, where $r_j$ is any value in $\eff{}{q}$.

\item For all $\hat{\vct{y}}_1, \hat{\vct{y}}_2 \in \eff{n}{q}$, $\bar{\beta}_1, \bar{\beta}_2, \vct{h} \in \eff{m}{q}$, and $\alpha_1, \alpha_2 \in \eff{}{q}$, let $\hat{\vct{y}} = \alpha_1 \hat{\vct{y}}_1 + \alpha_2 \hat{\vct{y}}_2$ and $\bar{\beta} = \alpha_1 \bar{\beta}_1 + \alpha_2 \bar{\beta}_2$. We require that if $\test(\pp, \hat{\vct{y}}_i, \bar{\beta}_i, \vct{h}) = \top$ for $ i = 1,2$ then $\test(\pp, \hat{\vct{y}}, \bar{\beta}, \vct{h}) = \top$.
\squishend

{\flushleft \bf Security.}\quad Let $\mathcal{H} = (\setup, \hash, \test)$ be a homomorphic hash. Let $\mathcal{A}$ be a probabilistic polynomial time (PPT) adversary that takes as input $\pp \leftarrow \setup (1^\lambda,n)$ and outputs $\vct{v}^* \in \eff{n+m}{q}$, an $m$-dimensional space $V$ represented as basis vectors $\hat{\vct{v}}_1, \cdots, \hat{\vct{v}}_m$, and a hash vector $\vct{h} \in \eff{m}{q}$.

\begin{definition}\label{defn:breakHash}
We say that $\mathcal{A}$ breaks the homomorphic hash scheme $\mathcal{H}$ if (i) $\hat{\vct{v}}^* \notin V$, (ii) $\test(\pp,\hat{\vct{v}}_i, \vct{e}_i, \vct{h}) = \top$ for $i=1,\cdots,m$, and (iii) $\test(\pp,\hat{\vct{v}}^*, \aug{v^*}, \vct{h}) = \top$. We define the {\em advantage} {\em Hash-Adv[}$\mathcal{A}$, $\mathcal{H}$\emph{]} of $\mathcal{A}$ to be the probability that $\mathcal{A}$ breaks $\mathcal{H}$. We say that $\mathcal{H}$ is {\em secure} if for all PPT $\mathcal{A}$,  {\em Hash-Adv[}$\mathcal{A}$, $\mathcal{H}$\emph{]} is negligible in the security parameter $\lambda$.
\end{definition}

{\flushleft {\bf Example Homomorphic Hash} $\mathcal{H}$-{\bf DL.}}\quad This construction is based on $\mathcal{VH}$-DL \cite{Agrawal2010} but customized to work with our augmentation scheme.
\squishlist
\item $\setup(1^\lambda, n)$:\\
-- Choose a finite cyclic group $\mathbb{G}$ of prime order $q > 2^\lambda$.\\
-- Choose generators $g_i \overset{\text{R}}{\leftarrow} \mathbb{G} \setminus \{1\}$ for $i=1, \cdots, n$.\\
-- Output $\pp := q, (g_1, \cdots, g_n)$ and the description of $\mathbb{G}$.

\item $\hash(\pp, \hat{\vct{v}})$:\\ 
-- Output $h := \prod_{i=1}^{n} \exp(g_i, {\vct{v}^{(i)}})$, where $\exp(a,b) = a^b$.

\item $\test(\pp, \hat{\vct{y}}, \bar{\beta}, \vct{h})$: If 
\[\prod_{i=1}^n \exp(g_i, {\vct{y}^{(i)}}) = \prod_{i=1}^m \exp(\vct{h}^{(i)}, \bar{\beta}^{(i)})\]
then output $\top$; otherwise, output $\bot$.
\squishend

The correctness conditions hold as follows:
\begin{enumerate}[(1)]
\item For all $\hat{\vct{v}} \in \eff{n}{q}$, if $h = \hash(\pp, \hat{\vct{v}})$ then for all $j \in [1,m]$:
\begin{align*}
\prod_{i=1}^n \exp(g_i, {\vct{v}^{(i)}}) &= h\\
\prod_{i=1}^m \exp(\vct{h}^{(i)}, \vct{e}_j^{(i)}) &=  h^1 \prod_{i=1,i \neq j}^m r_j^0 = h
\end{align*}
As a result, $\test(\pp, \hat{\vct{v}}, \vct{e}_j, \vct{h}) = \top\,$.

\item For all $\hat{\vct{y}}_1, \hat{\vct{y}}_2 \in \eff{n}{q}$, $\bar{\beta}_1, \bar{\beta}_2, \vct{h} \in \eff{m}{q}$, and $\alpha_1, \alpha_2 \in \eff{}{q}$, let $\hat{\vct{y}} = \alpha_1 \hat{\vct{y}}_1 + \alpha_2 \hat{\vct{y}}_2$ and $\bar{\beta} = \alpha_1 \bar{\beta}_1 + \alpha_2 \bar{\beta}_2$. If $\test(\pp, \hat{\vct{y}}_i, \bar{\beta}_i, \vct{h}) = \top$ for $ i = 1,2$ then 
\begin{align*}
~& \prod_{i=1}^n \exp(g_i, {\vct{y}^{(i)}}) = \prod_{i=1}^n \exp(g_i, {   \alpha_1 \hat{\vct{y}}_1^{(i)}  + \alpha_2 \hat{\vct{y}}_2^{(i)}  })\\
~&= \left[\prod_{i=1}^n \exp(g_i, {   \hat{\vct{y}}_1^{(i)} }) \right]^{\alpha_1} \left[\prod_{i=1}^n \exp(g_i, {   \hat{\vct{y}}_2^{(i)} })\right]^{\alpha_2}\\
~&= \left[\prod_{i=1}^m \exp(\vct{h}^{(i)}, \bar{\beta}_1^{(i)}  )\right]^{\alpha_1} \left[\prod_{i=1}^m \exp(\vct{h}^{(i)}, \bar{\beta}_2^{(i)}  )\right]^{\alpha_2}\\
~&= \prod_{i=1}^m \exp(\vct{h}^{(i)}, \alpha_1 \bar{\beta}_1^{(i)} + \alpha_2 \bar{\beta}_2^{(i)} )\\
~&= \prod_{i=1}^m \exp(\vct{h}^{(i)}, \bar{\beta})\,.
\end{align*}
As a result, $\test(\pp, \hat{\vct{y}}, \bar{\beta}, \vct{h}) = \top$.
\end{enumerate}

\begin{theorem}\label{thm:secureHash}
The homomorphic hash {\em \HDL} is secure assuming the discrete logarithm problem in $\mathbb{G}$ is hard. In particular, let $\mathcal{A}$ be a PPT adversary that breaks {\em \HDL}, then there exists a polynomial-time algorithm $\mathcal{B}$ that computes discrete logarithms in $\mathbb{G}$ such that {\em Hash-Adv[$\mathcal{A}$, \HDL] $\leq$ 2 $\cdot$ DL-Adv[$\mathcal{B}$, $\mathbb{G}$]}, where {\em DL-Adv[$\mathcal{B}$, $\mathbb{G}$]} is the probability that $\mathcal{B}$ computes discrete logarithms in $\mathbb{G}$ (formally defined in \cite{Katz2007}).
\end{theorem}

\begin{IEEEproof}[Proof (based on proof of {\em $\mathcal{VH}$-DL)}]
If $\mathcal{A}$ can break \HDL, he can output $\vct{v}^*, \hat{\vct{v}}_1, \cdots, \hat{\vct{v}}_m$, and $\vct{h}$ that satisfy definition \ref{defn:breakHash}. Thus, 
\[\prod_{i=1}^n \exp(g_i, {\hat{\vct{v}}^{*(i)}}) = \prod_{i=1}^m \exp(\vct{h}^{(i)}, \aug{\vct{v}^*}^{(i)})\] 
Let $ \hat{\vct{v}} = \sum_{i=1}^{m} \aug{\vct{v}^*}^{(i)}\,\hat{\vct{v}}_i $. Since $\hat{\vct{v}}^*$ is not a linear combination of $\hat{\vct{v}}_1, \cdots, \hat{\vct{v}}_m$, $\hat{\vct{v}}^* \neq \hat{\vct{v}}$. Since $\test(\pp,\hat{\vct{v}}_i, \vct{e}_i, \vct{h}) = \top$ for $i=1,\cdots,m$, and $\hat{\vct{v}}$ is a linear combination of $\hat{\vct{v}}_1, \cdots, \hat{\vct{v}}_m$, $\test(\pp,\hat{\vct{v}}, \aug{\vct{v}^*} , \vct{h}) = \top$. This means
\[\prod_{i=1}^n \exp(g_i, {\hat{\vct{v}}^{(i)}}) = \prod_{i=1}^m \exp(\vct{h}^{(i)}, \aug{\vct{v}^*}^{(i)})\] 
Consequently, $\mathcal{A}$ can find two distinct vector $\hat{\vct{v}}^*, \hat{\vct{v}} \in \eff{n}{q}$ such that 
\[\prod_{i=1}^n \exp(g_i, {\hat{\vct{v}}^{*(i)}}) = \prod_{i=1}^n \exp(g_i, {\hat{\vct{v}}^{(i)}})\,. \]
Assume $\mathcal{A}$ can find this collision with probability $\epsilon$ then $\mathcal{A}$ can be used to compute discrete logarithms in $\mathbb{G}$ with probability at least $\epsilon/2$ based on Theorem 3.4 in \cite{Bellare1994}.
\end{IEEEproof}

\subsection{Detection Guarantees}

Using the downloaded hashes, all nodes in the network can verify the integrity of all downloaded packets on-the-fly. The following theorem summarizes the security guarantee of our hash-based detection scheme.

\begin{theorem}\label{thm:hashDetection}
If a secure homomorphic hash scheme and a secure traditional hash function is used in the detection scheme, then the probability of a benign node accepting a corrupted packet is negligible in the security parameter. 
\end{theorem}

\begin{IEEEproof}
For a received packet $\vct{y}$, for nodes that are specified to perform decoding but cannot recover any source packets or nodes that are not specified to perform decoding, they verify the integrity of $\vct{y}$ using the verification of the homomorphic hash scheme. Let $\vct{h} = \{h_1, \cdots, h_m\}$, where $h_i, i \in [1,m]$, denotes the hash value of the data, $\hat{\vct{v}}_i$, of the source packet $\vct{v}_i$. A corrupted packet is a packet whose data is not in the committed source data space; hence, if $\vct{y}$ is corrupted then $\hat{\vct{y}} \notin \lspan{\hat{\vct{v}}_1, \cdots, \hat{\vct{v}}_m}$. As a result, the probability that any node $N$ in the network accepts a corrupted packet is upper bounded by the probability of breaking the homomorphic hash scheme, which is negligible in the security parameter $\lambda$. 

For a node $N$ that is specified to perform decoding and can recover a source packet from the decoding using $\vct{y}$ and previously (verified) received packets, it checks the integrity of $\vct{y}$ through checking the integrity of the newly recovered source packet. The probability of accepting a corrupted $\vct{y}$ is now dependent not only on the probability that the newly recovered source packet is corrupted but passing the verification but also on the probability that some of the previously received packets are corrupted but passed the verification. The proof is by induction:

Let $\negl$ denote a negligible function. Let $p_x$ and $p_y$ be the probabilities of breaking the traditional hash and homomorphic hash functions, respectively. Note that both of these probabilities are negligible. Let $\vct{y}_i$ denote packet $i$-th that arrives at node $N$. Let $ \text{Pr}[\vct{y}_i] $ denote the probability that node $N$ accepts a corrupted packet $\vct{y}_i$. The first packet is either a source packet or not, thus $N$ performs either a traditional hash check or homomorphic hash check. Hence,
\[  \text{Pr}[\vct{y}_1] = p_x + p_y = \negl \]
If the $t$-th packet is decodable, let
$\vct{y}_t = \sum_{i=1}^{t-1} \alpha_i \vct{y}_i + \beta \vct{v} \,,$
where $\vct{v}$ is the newly recovered source packet; $\vct{y}_i$'s are previously received, verified packets; $\alpha_i$'s and $\beta$ are some integer coefficients. The probability that $\vct{y}_t$ is corrupted but accepted by $N$, is
\begin{align*}
\text{Pr}[\vct{y}_t] &= \sum_{i=1}^{t-1} \text{Pr}[\vct{y}_i] \cdot \text{Pr}[\alpha_i \neq 0] + p_x + p_y\\
~ &\leq  \sum_{i=1}^{t-1} \text{Pr}[\vct{y}_i] + \negl
\end{align*}
Since $t$ is upper bounded by $c\,m$, where $c$ is some small positive integer, and $\text{Pr}[\vct{y}_1] = \negl$, $\text{Pr}[\vct{y}_t]$ is negligible for all $t \leq c\,m$.
\end{IEEEproof}

Finally, our hash-based detection scheme is collusion resistant because collusion does not help to break the discrete log assumption or a secure traditional hash function.

\section{The MAC-Based Defense}
\label{sec:mac_based}


\subsection{Key Observation}
{\flushleft \bf Observation 3.}\quad Let us revisit the discussion of applying homomorphic MAC scheme in Section \ref{subsec:threat_models}. From the attack, we observe that it is necessary that (i) each source generates tags of its packets using its own secret key as opposed to using a common key, or (ii) the controller generates all the tags under a key secret to all the sources.


\subsection{Homomorphic Multi-Source MAC ($\IMac$)}
\label{subsec:InterMac} 

In this section, we present a novel multi-source homomorphic MAC scheme, called $\IMac$, that allows different sources to generate tags using different keys. Nonetheless, the tags are combinable, and the malicious nodes cannot generate valid tags of corrupted packets.

{\flushleft \em Definitions:}\quad A ($q,n,s,g$) multi-source homomorphic MAC scheme is defined by four PPT algorithms:
\squishlist
\item $\gnr(\iden, k, \Pi)$: Input: a source space/generation identifier, $\iden$; a secret key, $k \in \bar{\mathcal{K}}$, and a committed source space, $\Pi$. $k$ is only known to the trusted controller and used for bootstrapping the MAC keys. Output: a key set $\mathcal{K} \triangleq \{k_1, \cdots, k_s\}$.
The $\iden$ is the unique source space/generation identifier. Given the committed source space $\Pi$, the $\gnr$ algorithm generates $s$ keys, where the $i$-th key can be used by source $i$ to generate tags for its source packets.

\item $\sign(k_i,\vct{v})$: Input: key $k_i \in \mathcal{K}$ used by source $S_i$ and a source packet $\vct{v}$ sent by source $S_i$. Output: tag $t$ of $\vct{v}$.
Let $\vct{v}_{i,1}, \cdots, \vct{v}_{i,g}$ denote source packets sent by source $S_i$. The $\sign$ algorithm signs the source space, $\Pi$, spanned by the source packets of all the sources by running $\sign(\iden, k_i, \vct{v}_{i,j})$, for all $i \in [1,s], j \in [1,g]$.

\item $\cbn((\mathbf{y}_{1},t_{1},\alpha_{1}), \cdots, (\mathbf{y}_{\ell},t_{\ell},\alpha_{\ell}))$: Input: $\ell$ ($\ell > 0$) vectors $\vct{y}_1, \cdots, \vct{y}_\ell \in \mathbb{F}^{n+m}_q$; their tags $t_1, \cdots, t_\ell \in \mathbb{F}_q$; and their coefficients $\alpha_1, \cdots, \alpha_\ell \in \mathbb{F}_q$. Output: tag $t$ of vector $\vct{y} \triangleq \sum_{i=1}^\ell \alpha_i\,\vct{y}_i$.

\item $\vrf(\mathcal{K}, \vct{y}, t)$: Input: a key set $\mathcal{K}$, a vector $\mathbf{y} \in \mathbb{F}^{n+m}_q$, and its tag $t \in \mathbb{F}_q$. Output: 0 (reject) or 1 (accept).
\squishend

{\flushleft \em Correctness:}\quad The scheme must satisfy the following correctness requirement: Let $\Pi$ be the committed source space spanned by the committed source packets of all the sources: $\vct{v}_{i,j}$, for all $i \in [1,s]$ and $j \in [1,g]$. Let $\Pi$'s identifier be $\iden$. Let $k \in \bar{\mathcal{K}}$, and $\mathcal{K} = \{k_1, \cdots, k_s\}$ be the output of $\gnr$ given $\iden$, $k$, and $\Pi$. Let $t_{i,j} = \sign(k_i,\vct{v}_{i,j})$ and $\alpha_{i,j} \in \eff{}{q}$, for all $i \in [1,s]$ and $j \in [1,g]$. Let $t = \cbn((\mathbf{v}_{1,1},t_{1,1},\alpha_{1,1}), \cdots, (\mathbf{v}_{s,g},t_{s,g},\alpha_{s,g}))$. Then 
\[\vrf \left( \mathcal{K}, \, \sum_{i=1}^{s} \sum_{j=1}^{g} \alpha_{i,j} \, \vct{v}_{i,j}, \, t \right) = 1\,.\]

{\flushleft \em Security:}\quad
We define the security using the following game:

{\flushleft \bf Attack Game.}\quad We consider the following attack game for a multi-source homomorphic MAC $\mathcal{T}$ = ($\gnr$, $\mac$, $\cbn$, $\vrf$), a challenger $\mathcal{C}$, and an adversary $\mathcal{A}$:

\squishlist
\item \emph{Setup:} The challenger generates a random key $k \overset{R}{\leftarrow} \bar{\mathcal{K}}$.

\item \emph{Queries:} $\mathcal{A}$ adaptively queries $\mathcal{C}$. Each query is of the form $(\iden_l, \Pi_l)$, where $\Pi_l$ is a linear subspace represented by a basis of $m$ vectors, $\vct{v}_{i,j}, i \in [1,s], j \in [1,g]$, and $\iden_l$ is the space identifier. We require that all identifiers $\iden_l$ submitted by $\mathcal{A}$ are distinct. To respond to a query for $(\iden_l, \Pi_l)$, the challenger does the following: Run $\gnr(\iden_l, k, \Pi_l)$ to produce a key set $\mathcal{K}_l = \{k_1, \cdots, k_s\}$. Compute $t_{i,j} = \sign(k_i, \vct{v}_{i,j})$, for all $i \in [1,s], j \in [1,g]$. Send ($t_{1,1}, \cdots, t_{s,g}$) and all keys in $\mathcal{K}_l$ {\em but one} to $\mathcal{A}$.

\item \emph{Output:} The adversary $\mathcal{A}$ outputs a triplet ($\iden_*, \vct{y}_*, t_*$). We consider that the adversary wins the security game if \\
(i) $\iden_* = \iden_l$ for some $l$,\\
(ii) $\vct{y}_* \notin \Pi_l$, and\\
(iii) $\vrf (\mathcal{K}_l, \vct{y}_*, t_*) = 1$.
\squishend

Requirement (i) is necessary as corrupted packet is only defined when there is a committed source space. Requirement (ii) indicates that the output packet by $\mathcal{A}$ is indeed a corrupted packet. Finally, (iii) indicates that $\mathcal{A}$ successfully forges a valid tag of the corrupted packet. Let Adv[$\mathcal{A}, \mathcal{T}$] denote the probability that $\mathcal{A}$ wins the above attack game. We define a secure multi-source homomorphic MAC scheme as follows:

\begin{definition}\label{defn:multiSrc}
A (q, n, s, g) multi-source homomorphic MAC scheme $\mathcal{T}$ is secure if and only if for all PPT adversaries $\mathcal{A}$, \emph{Adv[$\mathcal{A}, \mathcal{T}$]} is negligible.
\end{definition}

{\flushleft \bf The Construction of $\IMac$.}\quad We now present our construction of $\IMac$. The key ingredient of this construction is the generation of the key set $\mathcal{K}$ so that each source can compute tags of its source packets using its own key; nonetheless, the tags are still combinable.

\squishlist 
\item $\gnr(\iden, k, \Pi)$:\\
-- Let $\vct{v}_{1,1}, \cdots, \vct{v}_{s,g} \in \eff{n+m}{q}$ be the committed source packets that span $\Pi$, and let them be represented as row vectors. For each $p \in [1,s]$,  let $M_p$ be a matrix whose rows are vectors in the following set
\[\{\vct{v}_{i,j} \, | \, i = 1, \cdots, s; \, i \neq p; \, j = 1, \cdots, g\} \,.\]
In other words, $M_p$ is a matrix consisted of committed source packets of all other sources but source $S_p$. Note that $\mathsf{rank}(M_p) = m-g$. Let $\Pi_{M_p}$ denote the space spanned by the rows of $M_p$.\\
-- The null space of the matrix $M_p$, denoted as $\Pi^\bot_{M_p}$, is the set of all row vectors $\vct{z} \in \eff{n+m}{q} $ for which $M_p \, \vct{z}^\text{T} = \vct{0}$. For any $(m-g) \times (n+m)$ matrix $M_p$, we have
\[ \mathsf{rank}(M_p) + \mathsf{nullity}(M_p) = n+m\]
known as rank-nullity theorem, where $\mathsf{nullity}(M_p)$ is the dimension of  $\Pi^\bot_{M_p}$. Thus, 
\[ \mathsf{dim}(\Pi^\bot_{M_p}) = n+m - (m-g) = n+g\,. \]
-- Let $\vct{b}_1, \cdots, \vct{b}_{n+g} \in \eff{n+m}{q}$ be a basis of $\Pi^\bot_{M_p}$. This basis can be found by solving $M_p \, \vct{z}^\text{T} = \vct{0}$. Let $F$ be a Pseudo Random Function (PRF): $\bar{\mathcal{K}} \times (\mathcal{I} \times [1,s] \times [1,n+g]) \rightarrow \eff{}{q}$, where $\mathcal{I}$ denotes the domain of the source space identifier. To generate key $k_p$ for source $S_p$, the controller computes\\
\hspace*{3mm} $\circ$ $r_i \leftarrow F(k,\iden,p,i) \in \eff{}{q}$, $\forall i \in [1,n+g]$.\\
\hspace*{3mm} $\circ$ $k_p \leftarrow \sum_{i=1}^{n+g} r_i \, \vct{b}_i \in \eff{n+m}{q}$.\\
-- Output: a key set $\mathcal{K} \triangleq \{k_1, \cdots, k_s\}$, where each key, $k_p, p \in [1,s]$, is generated as above.

\item $\sign(k_i,\vct{v})$: Outputs $t \leftarrow k_i \cdot \vct{v} \in \eff{}{q}\,$.

\item $\cbn((\vct{y}_1,t_1,\alpha_1), \cdots, (\vct{y}_\ell,t_\ell,\alpha_\ell))$: Outputs the sum $t \leftarrow \sum_{i=1}^\ell \alpha_i \, t_i \in \eff{}{q}\,$.

\item $\vrf(\mathcal{K}, \vct{y}, t)$: Compute $t' = \vct{y} \cdot (k_1 + \cdots + k_s) $, where $k_i \in \mathcal{K}$. If $t=t'$, output 1; otherwise, output 0.
\squishend

{\flushleft \em Correctness:}\quad 
Recall from the correctness requirement that 
\[t = \sum_{i=1}^s \sum_{j=1}^g \alpha_{i,j} \, t_{i,j} = \sum_{i=1}^s \sum_{j=1}^g \alpha_{i,j} ( \vct{v}_{i,j} \cdot k_i)\,.\]
Also, $t'$ computed by the verification algorithm equals
\[\sum_{i=1}^s \sum_{j=1}^g \alpha_{i,j} \, \vct{v}_{i,j} \cdot (k_1 + \cdots + k_s) \overset{\text{(1)}}{=} \sum_{i=1}^s \sum_{j=1}^g \alpha_{i,j} ( \vct{v}_{i,j} \cdot k_i)\,.\]
Equality (1) is because by construction, for all $i \neq p$, $i \in [1,s]$, $p \in [1,s]$, and $j \in [1,g]$, $\vct{v}_{i,j} \cdot k_p = 0$. As computed, $t'=t$.

{\flushleft \em Security:}\quad We prove the security of $\IMac$ assuming $F$ is a secure PRF. For a PRF adversary $\mathcal{B}$, we let PRF-Adv[$\mathcal{B},F$] denote $\mathcal{B}$'s advantage in winning the PRF security game w.r.t. $F$. The definition of the PRF security game is provided in \cite{Katz2007}.

\begin{theorem}\label{thm:InterMac}
For any fixed q, n, s, g, $\IMac$ is a secure (q, n, s, g) multi-source homomorphic MAC, assuming F is a secure PRF. In particular, for every multi-source homomorphic MAC adversary $\mathcal{A}$, there is a PRF adversary $\mathcal{B}$ who has similar running time to $\mathcal{A}$, such that
\emph{\[\text{Adv}[\mathcal{A},\IMac] \leq \text{PRF-Adv}[\mathcal{B}, F] + \frac{1}{q}\,.\]}
\end{theorem}

\begin{IEEEproof} The proof is by using a sequence of games denoted as Game 0 and 1. Let $W_0$ and $W_1$ denote the events that $\mathcal{A}$ wins the multi-source homormophic MAC security in Game 0 and 1, respectively. Let Game 0 be identical to Attack Game 1. Hence,
\begin{align}\label{eq:W0}
\text{Pr}[W_0] = \text{Adv}[\mathcal{A}, \IMac]\,.
\end{align}
In Game 1, the PRF $F$ is replaced by a truly random function, \ie, to respond to the queries, the challenger computes $k_p = \sum_{i=1}^{n+g} r_i \, \vct{x}_i$, where $r_i \overset{R}{\leftarrow} \eff{}{q}$ instead of $r_i \leftarrow F(k,\iden,p,i)$. Everything else remains the same. Then, there exists a PRF adversary $\mathcal{B}$ such that
\begin{align}\label{eq:W0W1}
| \text{Pr}[W_0] - \text{Pr}[W_1] | = \text{PRF-Adv}[\mathcal{B}, F]\,.
\end{align}
The complete challenger in Game 1 works as follows:

{\flushleft $\bullet$ \em Queries:} $\mathcal{A}$ submits MAC queries $(\iden, \Pi)$, where $\Pi = \mathsf{span}(\vct{v}_{1,1}, \cdots, \vct{v}_{s,g})$. For each $p \in [1,s]$, $\mathcal{C}$ computes a basis of $\Pi_{M_p}^\bot$: $\vct{x}_1, \cdots, \vct{x}_{n+g}$. Then, in order to generate $k_p$, $\mathcal{C}$ does\\
\hspace*{1mm} -- $r_i \overset{R}{\leftarrow} \eff{}{q}$, $\forall i \in [1,n+g]$ .\\
\hspace*{1mm} -- $k_p \leftarrow \sum_{i=1}^{n+g} r_i \, \vct{x}_i \in \eff{n+m}{q}$ .\\
In other words, each $k_p$ is chosen uniformly at random from $\Pi_{M_p}^\bot$, a subspace of size $q^{n+g}$. The challenger $\mathcal{C}$ then computes tags for the committed source packets. For $i = 1, \cdots, s$ and $j = 1, \cdots, g$,\\
\hspace*{1mm} -- $t_{i,j} \leftarrow k_i \cdot \vct{v}_{i,j}$ .\\
Finally, $\mathcal{C}$ sends all the tags and all the keys but one to $\mathcal{A}$. Without loss of generality, assume that $\mathcal{C}$ keeps $k_1$ secret to $\mathcal{A}$.

{\flushleft $\bullet$ \em Output.} $\mathcal{A}$ eventually outputs a triplet $(\iden_*, \vct{y}_*, t_*)$. Assume that $\iden_* = \iden_l$, for some $l$. Let $\mathcal{K}_l = \{k_1, \cdots, k_s\}$ denote the key set generated for query $(\iden_l, \Pi_l)$. The adversary wins the game, \ie, event $W_1$ happens, if\\
\hspace*{1mm} -- $\vct{y}_* \notin \Pi_l$, and\\
\hspace*{1mm} -- $t_* = \vct{y}_* \cdot (k_1 + \cdots + k_s)$\\
Note that the adversary knows $k_2, \cdots, k_s$, therefore, if $\vct{y}_* \cdot k_1$ is known, the adversary will be able to forge a valid $t_*$. In what follows, we will show that $\vct{y}_* \cdot k_1$ is indistinguishable from a random value in $\eff{}{q}$. Let $\Pi_l = \mathsf{span}(\vct{v}_{1,1}, \cdots, \vct{v}_{s,g})$. Consider the following system of linear equations:
\begin{align*}
\vct{v}_{1,1} \cdot k_1 &= t_{1,1}\\ 
~&\cdots \\
\vct{v}_{1,g} \cdot k_1 &= t_{1,g}\\
\vct{v}_{2,1} \cdot k_1 &= 0\\ 
~&\cdots \\
\vct{v}_{s,g} \cdot k_1 &= 0\\ 
\vct{y}_* \cdot k_1 &= t_* - \vct{y}_* \cdot (k_2 + \cdots + k_s)
\end{align*}
The first $sg$ equations represent all information that the adversary learns about $k_1$ from its query $(\iden_l, \Pi_l)$. Note that since $\vct{y}_* \notin \Pi_l$, $\vct{y}_*$ and $\vct{v}_{i,j}$ ($i \in [1,s], j \in [1,g]$) are linearly independent. As a result, the above system of equations is consistent regardless of the value of $t_*$ because the coefficient matrix has rank $s g+1$ which equals the number of equations. Furthermore, for a fixed $\vct{y}_*$, for any value $t_* \in \eff{}{q}$, the solution space always has the same size $q^{n+sg-(sg+1)} = q^{n-1}$. Because $k_1$ is chosen uniformly at random from $\Pi_{M_1}^\bot$, and all solutions to the above system of equations are in $\Pi_{M_1}^\bot$, for a fixed $\vct{y}_*$, its valid tag $t_*$ could be any value in $\eff{}{q}$ equally likely. As a result, the probability that the adversary chooses a correct $t_*$ is $\frac{1}{q}$. Thus,
\begin{align}\label{eq:W1}
\text{Pr}[W_1] = \frac{1}{q}\,.
\end{align}
Equations (\ref{eq:W0}), (\ref{eq:W0W1}), and (\ref{eq:W1}) together prove the theorem.
\end{IEEEproof}

Theorem \ref{thm:InterMac} expresses that an adversary $\mathcal{A}$ can only forge a valid tag of a corrupted packet with probability $\frac{1}{q}$. This security guarantee may be unsatisfactory when working with a small field, \eg, $q=2^8$. Nevertheless, as suggested in \cite{Agrawal2009, Li2010, Zhang2011, Le2011}, the security can be improved by increasing the field size or using multiple tags. 
When using $\ell$ tags, the security is $\frac{1}{q^\ell}$. Note that using multiple tags to increase the security is preferable as increasing the field size increases the field multiplication complexity logarithmically \cite{Li2010}.

\begin{figure}
\centering
\begin{tikzpicture}
\node[scale=0.8] {
\begin{tikzpicture}
 	\tikzstyle{circ} = [circle,draw=black!50,fill=black!10,thick]
	\tikzstyle{elip} = [ellipse,draw=black!50,fill=blue!10,thick, minimum height=1.5cm,minimum width=2.5cm,]
	\tikzstyle{pre} = [<-,shorten <=1pt,>=stealth',semithick]
	\tikzstyle{post}=[->,shorten >=1pt,>=stealth',semithick]
	\node[elip] (network)	at (3,1.5)		{\Large Network};
	\node[circ]	(S1)		at (0,3)			{$S_1$}
		edge [post] (network);
	\node[circ]	(S2)		at (0,2)			{$S_2$}
		edge [post] (network);
	\node[circ]	(S3)		at (0,1)			{$S_3$}
		edge [post] (network);
	\node[circ]	(S4)		at (0,0)			{$S_4$}
		edge [post] (network);
	\node[circ]	(R1)		at (6,3)			{$R_1$}
		edge [pre] (network);
	\node[circ]	(R2)		at (6,2)			{$R_2$}
		edge [pre] (network);
	\node[circ]	(R3)		at (6,1)			{$R_3$}
		edge [pre] (network);
	\node[circ]	(R4)		at (6,0)			{$R_4$}
		edge [pre] (network);
	\node[left] 			at (S1.west)		{$k_1$};
	\node[left] 			at (S2.west)		{$k_2$};
	\node[left] 			at (S3.west)		{$k_3$};
	\node[left] 			at (S4.west)		{$k_4$};
	\node[right] 			at (R1.east)		{ $k_1+k_2+k_3$};
	\node[right] 			at (R2.east)		{ $k_1+k_2+k_4$};
	\node[right] 			at (R3.east)		{ $k_2+k_3+k_4$};
	\node[right] 			at (R4.east)		{ $k_1+k_2+k_3+k_4$};	
\end{tikzpicture}
};
\end{tikzpicture}
\caption{An example demonstrating the minimum amount of information required for carrying out the verification at each receiver when using $\IMac$.}
\label{fig:interMacKeys}
\end{figure}
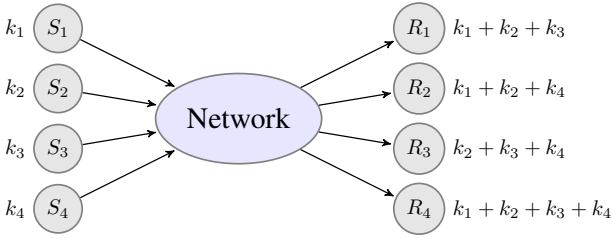

{\flushleft \bf Remarks.} We make the following two important observations w.r.t. the verification done in $\IMac$: (i) a node only needs to know the sum of the keys for the verification, and (ii) when there is an upper bound $M$ on the number of possible malicious sources, it may suffice for a verifying node to know the sum of just $M+1$ keys to carry out the verification.

For instance, consider the network given in Fig. \ref{fig:interMacKeys}. There are 4 source-receiver pairs: $(S_1, R_1), \cdots, (S_4, R_4)$. As discussed in Section \ref{subsec:interSession}, in inter-session network coding, a receiver does not always receive linear combination of source packets from all the sources. Assume that $R_1$ and $R_2$ only receive linear combinations of source packets sent by $S_1$ and $S_2$; $R_3$ only receives combinations of source packets sent by $S_2$, $S_3$; $S_4$ receives linear combinations of source packets from all the sources; and that the maximum number of malicious sources is 2. Then, the sum of the keys depicted at each receiver in Fig. \ref{fig:interMacKeys} is sufficient for each node to carry out the verification.

The reason why $(k_1 + k_2 + k_3)$ is sufficient for $R_1$ to verify a packet $\vct{y}$ is twofold: (i) If $\vct{y}$ is a benign packet, $k_4 \cdot \vct{y} = 0$ as $k_4 \in \Pi_{M_4}^\bot$; thus, $\vct{y} \cdot (k_1 + \cdots + k_4) = \vct{y} \cdot (k_1 + \cdots + k_3)$. As a result, $R_1$ does not need to know $k_4$ to verify a valid packet. (ii) If $\vct{y}$ is corrupted, since there is at least one key secret to the adversary ($M=2$), we can use the same line of arguments as in the proof of Theorem \ref{thm:InterMac} to show that the probability of forging a valid MAC tag for $\vct{y}$ is only $\frac{1}{q}$. 

Showing that the other sums are sufficient for $R_2$, $R_3$, and $R_4$ can be done with similar arguments. Having different sums for verification at different receivers decreases the damage done by the adversary who could compromise some of the receivers. We discuss this in detail at the end of Section \ref{subsec:privateMAC}.

\subsection{Efficient Commitment}
\label{subsec:commit}

The role of the committed source packets in $\IMac$ is to enable the controller to generate vectors (MAC keys) that are orthogonal to the committed space ($\Pi_{M_p}$'s). Here, we design a more efficient commitment scheme that does not require each source to send all their source packets to the controller, but it still allows the controller to generate these orthogonal vectors. To this end, we leverage two key techniques: padding for orthogonality and private inner product computation.

The padding for orthogonality technique was originally introduced in \cite{Zhang2011} to make a random vector orthogonal to all source packets of multiple generations by padding to each source packet an additional element. We apply this technique to make a random vector chosen by the controller, which will serve as a MAC key, orthogonal to the required subspace ($\Pi_{M_p}$). In addition, we use the private inner product protocol proposed in \cite{Goethals2004} to allow the controller to compute the padding elements while keeping the random chosen vector private.

{\flushleft \bf Private Inner Product Protocol.}\quad Let $\mathcal{E}$ = ($\mathsf{Gen}$, $\mathsf{Enc}$, $\mathsf{Dec}$) be a semantically secure homomorphic public-key cryptosystem. In general, the private inner product protocol (PIP) proposed in \cite{Goethals2004} works with various public-key cryptosystems that have the following homomorphic properties:
\begin{itemize}
\item $\mathsf{Dec}\left( \, \mathsf{Enc}(m_1) \, \mathsf{Enc}(m_2) \, \right) = m_1 + m_2$ , and
\item $\mathsf{Dec}\left( \, \mathsf{Enc}(m_1)^{m_2} \, \right) = m_1\,m_2$ .
\end{itemize}
Popular cryptosystems that possess the above properties include Goldwasser-Micali \cite{Goldwasser1984}, Paillier \cite{Paillier1999}, and Benaloh \cite{Benaloh1994} cryptosystems. However, not all of them are suited for our task. Specifically, in Paillier system, the plaintext must be in $\mathbb{Z}_q$, where $q$ is a product of two large primes, making $\mathbb{Z}_q$ not a finite field; this system thus does not fit our setting. In Goldwasser-Micali system, the plaintext domain is $\eff{}{2}$ and could be extended to $\eff{}{2^\ell}$ \cite{Franklin2010}; however, the {\em expansion factor}, \ie, the ratio between the size of the ciphertext and the plaintext, is very high (up to hundreds), making it not suitable for our purpose. Benaloh system is an extension of Goldwasser-Micali system: it reduces the expansion factor significantly; moreover, its plaintext domain could be a finite field $\mathbb{Z}_q$, where $q$ is prime. Therefore, we choose this system in our instantiation of the PIP protocol.

Let $q$ be prime, so that $\eff{}{q}$ is isomorphic to $\mathbb{Z}_q$. Let $\vct{r} = (r_1, \cdots, r_n)$ be a random vector chosen by the controller $C$, and $\vct{v} = (v_1, \cdots, v_n)$ be a source vector of source $S$. $C$ and $S$ carry out the PIP protocol described in Table \ref{table:innerProduct}. With PIP, $C$ can learn the inner product $\vct{r} \cdot \vct{v}$ while $S$ does not learn any information about $\vct{r}$, thanks to the security guarantee of the encryption. Indeed, Goethals \ea \cite{Goethals2004} showed that this protocol is secure in the {\em semi-honest model}, where it is assumed that both parties follow the protocol, but they are curious and try to deduce information from all exchanged data. 

\begin{table}[t]
\centering
{\normalsize
\begin{tabular}{|l|}
\hline
{\sc Private Inputs:} Private vectors $\vct{r}, \vct{v} \in \mathbb{F}^n_q$.\\
{\sc Private Outputs:} Inner product $\vct{r} \cdot \vct{v} \in \mathbb{F}_q$.\\
~\\
1. Setup phase. The controller $C$ does:\\
\quad\quad Generate a private and public key pair $(\mathsf{sk}, \mathsf{pk})$.\\
\quad\quad Send $\mathsf{pk}$ to $S$.\\
2. The controller $C$ does for $i \in [1,n]$:\\
\quad\quad Send $c_i = \mathsf{Enc}_{\mathsf{pk}}(r_i)$ to $S$.\\
3. The source $S$ does:\\
\quad\quad Send $w \leftarrow \prod_{i=1}^n c_i^{v_i}$ to $C$.\\
4. The source $S$ does:\\
\quad\quad Compute $\vct{r} \cdot \vct{v} = \mathsf{Dec}_{\mathsf{sk}}(w)$.\\
\hline
\end{tabular}
}
\caption{Private Inner Product (PIP) Protocol}
\label{table:innerProduct}
\end{table}

\begin{figure}[t]
\centering
\begin{tikzpicture}
\node[scale=0.7] {
\begin{tikzpicture}[bend angle=30]
\tikzstyle{circ} =[circle,draw=black!50,fill=black!10,thick]
\tikzstyle{circX} =[circle,draw=black!50,fill=green!40,thick]
\tikzstyle{pre} =[<-,shorten <=1pt,>=stealth',semithick,dashed,draw=black]
\tikzstyle{post}=[->,shorten >=1pt,>=stealth',semithick,dotted,draw=blue]
	\node[circX]	(C)					at (6,0)			{$C$};
	\node[circ]	(S1)					at (0,0)			{$S_1$}
		edge[post] node[midway,sloped,above,blue]		{$\mathsf{Enc}(\hat{\vct{r}}_2 \cdot \hat{\vct{v}}_1)$} (C)
		edge[pre, bend left] node[midway,sloped,above] 	{$\mathsf{Enc}(\hat{\vct{r}}_2)$} (C)
		edge[pre, bend right] node[midway,sloped,above] 	{$p_{1,1}^{(1)}$, $k_1=\vct{r}_1$} (C);
	\node[circ]	(S2)					at (12,0)			{$S_2$}
		edge[post] node[midway,sloped,above,blue]		{$\mathsf{Enc}(\hat{\vct{r}}_1 \cdot \hat{\vct{v}}_2)$} (C)
		edge[pre, bend right] node[midway,sloped,above] 	{$\mathsf{Enc}(\hat{\vct{r}}_1)$} (C)
		edge[pre, bend left] node[midway,sloped,above] 	{$p_{2,1}^{(1)}$, $k_2=\vct{r}_2$} (C);
	\node[above]				at (C.north)		{$\vct{r}_1, \vct{r}_2$};
\end{tikzpicture}
};
\end{tikzpicture}
\vspace*{-4mm}
\caption{Keys generation using the CPK protocol for the network of Fig. \ref{fig:multi-src-atk}. $k_1$ is orthogonal to the padded vector $\vct{p}_{2,1}$ and $k_2$ is orthogonal to $\vct{p}_{1,1}$ thanks to the padding. At the same time, $k_1$ is secret to $S_2$ and $k_2$ is secret to $S_1$.}
\label{fig:CPK}
\vspace*{-3mm}
\end{figure}
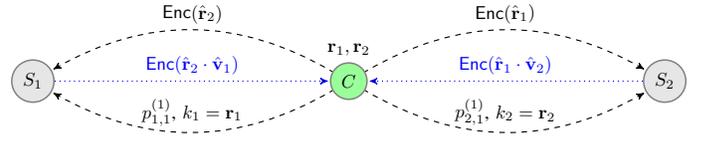

{\flushleft \bf Commitment, Padding, and Key Generation (CPK) Protocol.}\quad Let $k \in \bar{\mathcal{K}}$ and $F$ be a PRF: $\bar{\mathcal{K}} \times (\mathcal{I} \times [1,s] \times [n+s-1+m]) \rightarrow \eff{}{q}$. Each source packet will be padded with $s-1$ elements. Using PIP, the controller $C$ generates the MAC keys and computes the padding as follows:
\begin{enumerate}[1.]
\item {\em Setup:} Let $\iden$ be the subspace identifier. For $i \in [1,s]$ and $j \in [1,n+s-1+m]$, $C$ computes $r_i^{(j)} \leftarrow F(k,\iden,i,j)$. Let $\vct{r}_i = (r_i^{(1)}, \cdots, r_i^{(n+s-1+m)})$ and $\hat{\vct{r}}_i = (r_i^{(1)}, \cdots, r_i^{(n)})$. 

\item {\em Commitment:} For each $i \in [1,s]$, $C$ and $S_{i'}, i' \in [1,s] \setminus \{i\}$, carry out the PIP protocol so that $C$ learns $\hat{\vct{r}}_i \cdot \hat{\vct{v}}_{i',j}$, $\forall j \in [1,g]$. The encryption of these dot products sent from each source to the controller in the PIP protocol represent the commitment made by the sources.

\item {\em Padding:} Let $p_{i,j}^{(1)}, \cdots, p_{i,j}^{(s-1)}$ denote the padding elements for a source packet $\vct{v}_{i,j}$ sent by source $S_i$. The padded source packet, denoted by $\vct{p}_{i,j}$, has the following form:
\[(\vct{\hat{v}}_{i,j}, p_{i,j}^{(1)}, \cdots, p_{i,j}^{(s-1)}, \underbrace{\overbrace{0, \cdots, 0, 1}^{g(i-1)+j}, 0, \cdots, 0}_m) \in \eff{n+s-1+m}{q} \]
The padding elements are computed by solving the following system of $s-1$ linear equations:
\vspace*{-2mm}
\begin{align}\label{eqn:orthogonal}
\{ \vct{r}_{i'} \cdot \vct{p}_{i,j} = 0 \}_{i' \in [1,s] \setminus \{i\}}
\end{align}
For $s>1$, this system has $s-1$ unknowns and consists of $s-1$ linearly independent equations. Therefore, there is a unique solution for $p_{i,j}^{(1)}, \cdots, p_{i,j}^{(s-1)}$. $C$ then sends the padding elements to $S_i$. $S_i$ now sends $\vct{p}_{i,j}$ instead of $\vct{v}_{i,j}$.

\item {\em MAC keys:} $C$ uses $\vct{r}_i$ as MAC key $k_i$. Equations in (\ref{eqn:orthogonal}) ensure that the chosen key $k_i$ is orthogonal to $\Pi_{M_i}$.
\end{enumerate}
When using the CPK protocol, the sources no longer need to send all of their source packets to the controller. Instead, they only need to send an encryption of the inner product for every source packet, thereby significantly reducing the communication cost. Fig. \ref{fig:CPK} illustrates the CPK protocol for the network shown in Fig. \ref{fig:multi-src-atk}. We use $\IMac_\text{CPK}$ to denote the $\IMac$ construction when using the CPK protocol to generate MAC keys instead of $\gnr$.

{\flushleft \bf Security.}\quad The security of $\IMac_\text{CPK}$ in the semi-honest model comes from the security of PIP and $\IMac$.

Let $\gnr_\text{CPK}$ denote a new generation algorithm that takes as input a key $k \in \bar{\mathcal{K}}$ and an identifier $\iden$, and generates MAC keys using the CPK protocol. Let $\IMac_\text{CPK}$ denote the new $\IMac$ construction with $\gnr_\text{CPK}$. Consider Attack Game 1, previously described in Section \ref{subsec:InterMac}, with the following modified query step:
\begin{itemize}
\item {\em Queries:} The adversary $\mathcal{A}$ chooses a subspace $\Pi$ and its identifier $\iden$, then sends $(\iden)$ to the challenger $\mathcal{C}$. $\mathcal{A}$ can make a polynomial number of queries. To response to a query $\iden_l$, $\mathcal{C}$ initiates the CPK protocol with $\mathcal{A}$ to computes the MAC key set $\mathcal{K}_l$. Let $p_{i,j}^{(1)}, \cdots, p_{i,j}^{(s-1)}$ denote the padding elements of the source packet $\vct{v}_{i,j}$ sent by source $S_i$. For each padded source vector $\vct{p}_{i,j}$, $\mathcal{C}$ can also compute its MAC tag under key $k_i = \vct{r}_i$:
\begin{align*} 
t_{i,j} &= \vct{r}_i \cdot \vct{p}_{i,j} \\
&= \hat{\vct{r}}_i \cdot \hat{\vct{v}}_{i,j} + p_{i,j}^{(1)} r_i^{(n+1)} + \cdots + p_{i,j}^{(s-1)} r_i^{(n+s-1)} \\
~&\quad+ r_i^{(n+s-1+g(i-1)+j)}\,.
\end{align*}
Finally, $\mathcal{C}$ sends all the tags and all the MAC keys in $\mathcal{K}_l$ but one to $\mathcal{A}$.
\end{itemize}
The setup step, output step, and the winning conditions remain the same. The definition of security for multi-source homomorphic MAC is now with respect to the above modified attack game. Let Enc-Adv$[\mathcal{B}_2, \mathcal{E}]$ be the probability that $\mathcal{B}_2$ has more than a random guess to output the correct bit of the public-key encryption security experiment $\mathsf{PubK}^\mathsf{eav}_\mathcal{E}$. We refer the reader to \cite{Katz2007} for the experiment.

\begin{theorem}
For any fixed q, n, s, g, $\IMac_\text{\em CPK}$ is a secure (q, n, s, g) multi-source homomorphic MAC in the semi-honest model, assuming F is a secure PRF and $\mathcal{E}$ is a semantically secure public-key encryption. In particular, for every multi-source homomorphic MAC adversary $\mathcal{A}$, there is a PRF adversary $\mathcal{B}_1$ and a public-key encryption adversary $\mathcal{B}_2$ who have similar running time to $\mathcal{A}$, such that
\emph{\[\text{Adv}[\mathcal{A},\IMac_\text{CPK}] \leq \text{PRF-Adv}[\mathcal{B}_1, F] + \text{Enc-Adv}[\mathcal{B}_2, \mathcal{E}] + \frac{1}{q}\,.\]}
\end{theorem}

\begin{IEEEproof}[Proof]
The proof is by using a sequence of games denoted as Game 0, 1, and 2. Let $W_0$, $W_1$ and $W_2$ denote the events that $\mathcal{A}$ wins the multi-source homormophic MAC security in Game 0, 1, and 2, respectively. Let Game 0 be identical to the modified Attack Game 0. Hence,
\begin{align}\label{eq:W0a}
\text{Pr}[W_0] = \text{Adv}[\mathcal{A}, \IMac_\text{CPK}]\,.
\end{align}

In Game 1, the PRF $F$ is replaced by a truly random function, \ie, in the CPK setup, the challenger computes $r^{(j)}_i \overset{R}{\leftarrow} \eff{}{q}$ instead of $r_i^{(j)} \leftarrow F(k,\iden,i,j)$. Everything else remains the same. Then, there exists a PRF adversary $\mathcal{B}_1$ such that
\begin{align}\label{eq:W0W1a}
| \text{Pr}[W_0] - \text{Pr}[W_1] | = \text{PRF-Adv}[\mathcal{B}_1, F]\,.
\end{align}

In Game 2, the encryption $\mathcal{E}$ is replaced with a perfect encryption scheme, \ie, the encryption is information-theoretically secure. There exists an encryption adversary $\mathcal{B}_2$ such that 
\begin{align}\label{eq:W1W2a}
| \text{Pr}[W_1] - \text{Pr}[W_2] | = \text{Enc-Adv}[\mathcal{B}_2, \mathcal{E}]\,.
\end{align}

Note that in Game 2, (i) in the semi-honest model, the adversary follow the CPK protocol; (ii) the encryptions sent from the challenger give no information about the random chosen vectors, $\vct{r}_i$'s, to the adversary; and (iii) $\vct{r}_i$'s are indistinguishable from vectors chosen uniformly at random from $\eff{n+s-1+m}{q}$. Following the same line of argument as in the proof of Theorem \ref{thm:InterMac} gives
\begin{align}\label{eq:W2a}
\text{Pr}[W_2] = \frac{1}{q}\,.
\end{align}

Equations (\ref{eq:W0a}), (\ref{eq:W0W1a}), (\ref{eq:W1W2a}), and (\ref{eq:W2a}) together prove the theorem.
\end{IEEEproof}

In a stronger threat model, where malicious sources may not follow the protocol, the security guarantee of $\IMac_\text{CPK}$ could still be achieved by adding appropriate controller's responses for malicious behaviors. Malicious behaviors of the sources are limited to (i) not sending a well-formed encryption back for each query of $C$, and (ii) not padding the source packets appropriately. For (i), the controller could exclude any source with this behavior from the source list and only calculate MAC keys for the remaining sources. For (ii), not-properly padded packets will be dropped with high probability as they are highly likely to be outside of the committed source space.

\subsection{Private Inner Product MAC}
\label{subsec:privateMAC}

$\IMac$ explores the first direction of Observation 3, which suggests different sources should use different keys. In this section, we explore the other direction, which suggests that all tags of the source packets be generated by the trusted controller instead of the sources, and the MAC key be secret to the sources. In particular, we show how the PIP protocol could be combined with $\SMac$ previously proposed for intra-session network coding \cite{Le2010} to provide an alternative MAC-based scheme for detecting corrupted packets.

$\SMac$ consists of a triplet of algorithms: $\mac$, $\cbn$, and $\vrf$. The construction of $\SMac$ uses a PRF $F: \bar{\mathcal{K}} \times (\mathcal{I} \times [1,n+m]) \rightarrow \mathbb{F}_q$ and is as follows:

\squishlist
\item $\mac(k, \iden, \vct{y})$:  The tag $t \in \eff{}{q}$ of an input vector $\vct{y} \in \eff{n+m}{q}$ is computed by the following steps:\\
-- $\vct{r}  \leftarrow (F (k,  \iden, 1), \cdots, F (k,  \iden, n+m))$ .\\
-- $t \leftarrow \vct{y} \cdot \vct{r} \in \mathbb{F}_q$ .

\item $\cbn((\mathbf{y}_1,t_1,\alpha_1), \cdots, (\mathbf{y}_\ell,t_\ell,\alpha_\ell))$: The tag $t \in \eff{}{q}$ of $\vct{y} \triangleq \sum_{i=1}^\ell \alpha_i \, \vct{y}_i \in \eff{n+m}{q}$ is computed as follows:\\
-- $t \leftarrow \sum_{i=1}^\ell \alpha_i \, t_i \in \mathbb{F}_q$ .

\item $\vrf(k, \iden, \vct{y}, t)$: To verify if $t$ is a valid tag of $\vct{y}$ using key $k$, we do the following:\\
-- $\vct{r}  \leftarrow (F (k,  \iden, 1), \cdots, F (k,  \iden, n+m))$ .\\
-- $t' \leftarrow \vct{y} \cdot \vct{r}$ .\\
-- If $t' = t$, output 1 (accept); otherwise, output 0 (reject).
\squishend

{\flushleft \bf Private MAC (PM) Protocol.}\quad The controller and the sources carry the PM protocol to compute tags of the source packets. The PM protocol consists of the following steps:

\begin{enumerate}[1.]
\item {\em Setup:} Let $\iden$ be the current subspace identifier. $C$ computes $r^{(i)}  \leftarrow (F (k,  \iden, i), \forall i \in [1,n+m]$. Let $\hat{\vct{r}} = (r^{(1)}, \cdots, r^{(n)})$ and $\vct{r} = (r^{(1)}, \cdots, r^{(n+m)})$.

\item {\em Commitment:} For each $i \in [1,s]$, $C$ and $S_{i}$ carry out the PIP protocol that allows $C$ to learn $\hat{\vct{r}} \cdot \hat{\vct{v}}_{i,j}, \forall j \in [1,g]$. The encryption of the inner products sent by the sources to the controller are the commitment.

\item {\em MAC tags:} For $\vct{v}_{i,j}$, $C$ computes its tag $t_{i,j} = \vct{r} \cdot \vct{v}_{i,j} = \hat{\vct{r}} \cdot \hat{\vct{v}}_{i,j} + r^{(g(i-1)+j)}$.
\end{enumerate}

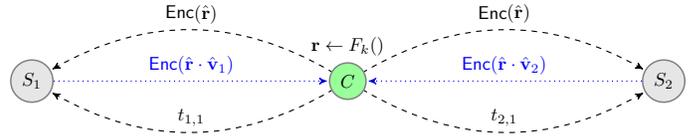
\begin{figure}[t]
\centering
\begin{tikzpicture}
\node[scale=0.7] {
\begin{tikzpicture}[bend angle=30]
\tikzstyle{circ} =[circle,draw=black!50,fill=black!10,thick]
\tikzstyle{circX} =[circle,draw=black!50,fill=green!40,thick]
\tikzstyle{pre} =[<-,shorten <=1pt,>=stealth',semithick,dashed,draw=black]
\tikzstyle{post}=[->,shorten >=1pt,>=stealth',semithick,dotted,draw=blue]
	\node[circX]	(C)					at (6, 0)			{$C$};
	\node[circ]	(S1)					at (0,0)			{$S_1$}
		edge[post] node[midway,sloped,above,blue]		{$\mathsf{Enc}(\hat{\vct{r}} \cdot \hat{\vct{v}}_1)$} (C)
		edge[pre, bend left] node[midway,sloped,above] 	{$\mathsf{Enc}(\hat{\vct{r}})$} (C)
		edge[pre, bend right] node[midway,sloped,above] 	{$t_{1,1}$} (C);
	\node[circ]	(S2)					at (12,0)			{$S_2$}
		edge[post] node[midway,sloped,above,blue]		{$\mathsf{Enc}(\hat{\vct{r}} \cdot \hat{\vct{v}}_2)$} (C)
		edge[pre, bend right] node[midway,sloped,above] 	{$\mathsf{Enc}(\hat{\vct{r}})$} (C)
		edge[pre, bend left] node[midway,sloped,above] 	{$t_{2,1}$} (C);	
	\node[above]						at (C.north)		{$\vct{r} \leftarrow F_k()$};
\end{tikzpicture}
};
\end{tikzpicture}
\vspace*{-4mm}
\caption{Tags generation using the PM protocol for the network of Fig. \ref{fig:multi-src-atk}. The tags are generated by the controller and the key is secret to the source.}
\label{fig:PM}
\vspace*{-6mm}
\end{figure}

Note that PM helps the controller compute the tags on behalf of the sources without leaking the MAC key. Fig. \ref{fig:PM} illustrates how the $\SMac$ MAC tags are computed for the network shown in Fig. \ref{fig:multi-src-atk}. We use $\SMac_\text{PM}$ to denote the $\SMac$ scheme when used with the PM protocol to generate tags for the source packets as opposed to the $\mac$ algorithm.

{\flushleft \bf Security.}\quad The security of $\SMac_\text{PM}$ in the semi-honest model comes from the security of PIP and $\SMac$. Below, we analyze the security of $\SMac$ when used with the PM protocol.
{\flushleft \bf Attack Game 2.}\quad 
We consider the following attack game for a homomorphic MAC $\mathcal{T}$ = ($\mac$, $\cbn$, $\vrf$), a challenger $\mathcal{C}$, and an adversary $\mathcal{A}$:
\begin{itemize}
\item {\em Setup:} $\mathcal{C}$ generates a random key $k \overset{R}{\leftarrow} \bar{\mathcal{K}}$ .

\item \emph{Queries.}  The adversary $\mathcal{A}$ chooses a subspace $\Pi$ and its identifier $\iden$, then sends $(\iden)$ to the challenger $\mathcal{C}$. $\mathcal{A}$ can make a polynomial number of queries. To response to a query $\iden_l$, $C$ initiates the PM protocol to compute tags of all source packets. $C$ then sends all the tags to $\mathcal{A}$.

\item \emph{Output.} The adversary $\mathcal{A}$ outputs a triplet ($\iden_*, \vct{y}_*, t_*$). We consider that the adversary wins the security game if
  \begin{enumerate}[(i)]
  \item $\iden_* = \iden_l$ for some $l$,
  \item $\vct{y}_* \notin \Pi_l$, and
  \item $\vrf (k, \iden_l, \vct{y}_*, t_*) = 1$.
  \end{enumerate}
\end{itemize}

Let Adv[$\mathcal{A}, \mathcal{T}$] denote the probability that $\mathcal{A}$ wins the above attack game. We define a secure homomorphic MAC scheme as follows:

\begin{definition}\label{defn:singleSrc}
A (q, n, m) homomorphic MAC scheme $\mathcal{T}$ is secure if for all probabilistic polynomial-time adversaries $\mathcal{A}$, \emph{Adv[$\mathcal{A}, \mathcal{T}$]} is negligible.
\end{definition}

Let $\SMac_\text{PM}$ denote the $\SMac$ scheme when used with the PM protocol to generate tags for the source packets as opposed to the $\mac$ algorithm. The security of $\SMac_\text{PM}$ is given by the following theorem:

\begin{theorem}
For any fixed q, n, m, $\SMac_\text{\em PM}$ is a secure (q, n, m) homomorphic MAC in the semi-honest model, assuming F is a secure PRF and $\mathcal{E}$ is a semantically secure public-key encryption. In particular, for every homomorphic MAC adversary $\mathcal{A}$, there is a PRF adversary $\mathcal{B}_1$ and a public-key encryption adversary $\mathcal{B}_2$ who have similar running time to $\mathcal{A}$, such that
\emph{\[\text{Adv}[\mathcal{A}, \SMac_\text{PM}] \leq \text{PRF-Adv}[\mathcal{B}_1, F] + \text{Enc-Adv}[\mathcal{B}_2, \mathcal{E}] + \frac{1}{q}\,.\]}
\end{theorem}

\begin{IEEEproof}[Proof]
The proof is by using a sequence of games denoted as Game 0, 1, and 2. Let $W_0$, $W_1$ and $W_2$ denote the events that $\mathcal{A}$ wins the homormophic MAC security in Game 0, 1, and 2, respectively. Let Game 0 be identical to the Attack Game 2. Hence,
\begin{align}\label{eq:W0b}
\text{Pr}[W_0] = \text{Adv}[\mathcal{A}, \SMac_\text{PM}]\,.
\end{align}

In Game 1, the PRF $F$ is replaced by a truly random function, \ie, in the PM setup, the challenger computes $r^{(i)} \overset{R}{\leftarrow} \eff{}{q}$ instead of $r^{(i)} \leftarrow F(k,\iden,i)$. Everything else remains the same. Then, there exists a PRF adversary $\mathcal{B}_1$ such that
\begin{align}\label{eq:W0W1b}
| \text{Pr}[W_0] - \text{Pr}[W_1] | = \text{PRF-Adv}[\mathcal{B}_1, F]\,.
\end{align}

In Game 2, the encryption $\mathcal{E}$ is replaced with a perfect encryption scheme, \ie, the encryption is information-theoretically secure. There exists an encryption adversary $\mathcal{B}_2$ such that 
\begin{align}\label{eq:W1W2b}
| \text{Pr}[W_1] - \text{Pr}[W_2] | = \text{Enc-Adv}[\mathcal{B}_2, \mathcal{E}]\,.
\end{align}

Note that in Game 2, (i) in the semi-honest model, the adversary follow the PM protocol; (ii) the encryptions sent from the challenger give no information about the random chosen vector, $\vct{r}$, to the adversary; and (iii) $\vct{r}$ is indistinguishable from a vector chosen uniformly at random from $\eff{n+m}{q}$. Let $\vct{v}_1, \cdots, \vct{v}_m$ be the source packets that span $\Pi_l$ (recall that $\iden_* = \iden_l$ for some $l$). Consider the following system of $m+1$ equations:
\begin{align*}
\vct{r} \cdot \vct{v}_1 &= t_1\\
\cdots &~\\
\vct{r} \cdot \vct{v}_m &= t_m\\
\vct{r} \cdot \vct{y}_* &= t_*
\end{align*}
The adversary learns the first $m$  equations from its query, and it wins the security game if the last equation is valid and $\vct{y}_* \notin \Pi_l$. This system of equations is consistent regardless of the value of $t_*$ because the coefficient matrix has rank $m+1$, which equals the number of equations. Furthermore, for any value $t_*$, the solution space always has the same size $q^{n-1}$. Thus, for a fixed $\vct{y}^*$, its valid tag $t_*$ could be any value in $\mathbb{F}_q$ equally likely, given that $\vct{r}$ is chosen uniformly at random from $\eff{n+m}{q}$. As a result, the probability that the adversary chooses a correct $t_*$  for any $\vct{y}_*$ is $\frac{1}{q}$, \ie,
\begin{align}\label{eq:W2b}
\text{Pr}[W_2] = \frac{1}{q}\,.
\end{align}

Equations (\ref{eq:W0b}), (\ref{eq:W0W1b}), (\ref{eq:W1W2b}), and (\ref{eq:W2b}) together prove the theorem.
\end{IEEEproof}

We note that the security of $\SMac_\text{PM}$ can also be extended to the malicious model, where there are sources that may not follow the PM protocol. In this model, a malicious source $S_i$ is limited to not sending back an encryption (of the inner product of $\vct{r}$ and the appropriate $\vct{v}_{i,j}$) or sending back a mal-form encryption. In response to these behaviors, the controller can ignore $\vct{v}_{i,j}$ in its tag computation and thus, do not send the tag of  $\vct{v}_{i,j}$ back to $S_i$. The source $S_i$, without knowing the key, $k$, will not be able to generate a valid tag for $\vct{v}_{i,j}$ (unless $\vct{v}_{i,j}$ is a linear combination of vectors with already known tags).


{\flushleft \bf Comparison.} Compared to $\IMac_\text{CPK}$, $\SMac_\text{PM}$ is simpler in terms of initialization. This is because $\IMac_\text{CPK}$ operates on $s$ MAC keys instead of one key. $\IMac_\text{CPK}$ and $\SMac_\text{PM}$ have similar efficient $\cbn$ and $\vrf$ operations as both of them only involve simple field addition and multiplication as opposed to exponentiation. When using $\SMac_\text{PM}$, all receivers must know the MAC key $k$ in order to verify their received packets. As a result, as soon as an adversary compromises a receiver and learns $k$, it can fool all other receivers into accepting corrupted packets. We stress that this is not necessarily the case when using $\IMac_\text{CPK}$. For instance, consider Fig. \ref{fig:interMacKeys}. Assume that $S_1$ and $S_2$ are malicious, thus keys $k_1$ and $k_2$ are leaked. If the adversary compromises $R_1$, it learns $k_3$ by subtracting the sum $(k_1 + k_2 + k_3)$ from $(k_1 + k_2)$. However, it still cannot fool $R_2$, $R_3$, or $R_4$ into accepting a corrupted packet as the verification at these receivers involves $k_4$, which is still secret to the adversary. 

$\IMac_\text{CPK}$ and $\SMac_\text{PM}$, as described, could be used as a drop-in replacement for traditional MACs, \eg, \HMac, for networks that use inter-session network coding: they allow the receivers to detect corrupted packets. As when using a traditional MAC scheme, we assume the keys distribution is through secure (athentic and private) channels. We also assume the communication between the sources and the controllers in the CPK and PM protocols is through athentic channels. In fact, compromising any node but $R_4$ does not help the adversary to break the verification of any additional receiver, and compromising $R_4$ only allows the adversary to break the verification of one additional receiver, $R_3$, but not all. 

\subsection{In-Network Detection}
Both of our MAC schemes could be extended to provide in-network detection by adopting state-of-the-art techniques proposed for intra-session network coding. We discuss two main options below:

{\flushleft \em Delayed Key Disclosure (TESLA) \cite{Perrig2002}:} This approach leverages the time dimension to achieve broadcast authentication and has been adapted to intra-session network coding to provide in-network detection \cite{Dong2009, Li2010, Le2011a}. In this approach, nodes are required to loosely synchronize their time. Both $\IMac_\text{CPK}$ and $\SMac_\text{PM}$ could be used with the approaches proposed in \cite{Li2010} and \cite{Le2011a} to provide in-network detection for fixed directed acyclic networks and dynamic peer-to-peer networks, respectively. We note that the detection schemes based on \cite{Li2010, Le2011a} are fully collusion resistant and tag-pollution resistant (an attack on MAC-based schemes that use multiple tags \cite{Li2010}).

{\flushleft \em Cover-Free Set Systems \cite{Canetti1999}:} This approach leverages cover-free set systems to probabilistically distribute keys to all nodes such that any collusion of $c$ nodes or less does not leak all the keys used in the whole system. This approach has been adapted to intra-session network coding to provide in-network detection \cite{Agrawal2010, Zhang2011}. Both $\IMac_\text{CPK}$ and $\SMac_\text{PM}$ are suitable to be used with this approach. Detection schemes based on \cite{Agrawal2010, Zhang2011} are $c$-collusion resistant. To address tag pollution, we propose using our homomorphic hash-based detection scheme to protect the coding coefficients and the tags of the packets. This technique is motivated by the hybrid scheme $\mathsf{MacSig}$ proposed by Zhang \ea \cite{Zhang2011}, where a homomorphic signature scheme is used to protect the coding coefficients and the tags.

\section{Performance Evaluation}
\label{sec:evaluation}

\subsection{Bandwidth Overhead}
\label{subsec:bandwidthOverhead}
We compute the bandwidth overhead directly from the number of packets, hashes, and MAC tags described in our schemes.

\subsubsection{Hash-Based Detection} Our hash-based scheme does not incur any online bandwidth overhead per packet as there is no additional symbol attached to each packet. The off-line bandwidth overhead of this scheme is dominated by the bandwidth required to distribute both the homomorphic and traditional hashes from the controller to all the nodes. The size of a homomorphic hash is $|q|$. Let $|\bar{h}|$ denote the size of the traditional hash (for $\SHA$, $|\bar{h}|$=160 bits). The total off-line bandwidth overhead is $s g |\mathcal{G} | ( |\bar{h}| + |q| )$.

\subsubsection{MAC-Based Detection}  
\begin{figure}[t]
\centering
\includegraphics[width=8.5cm]{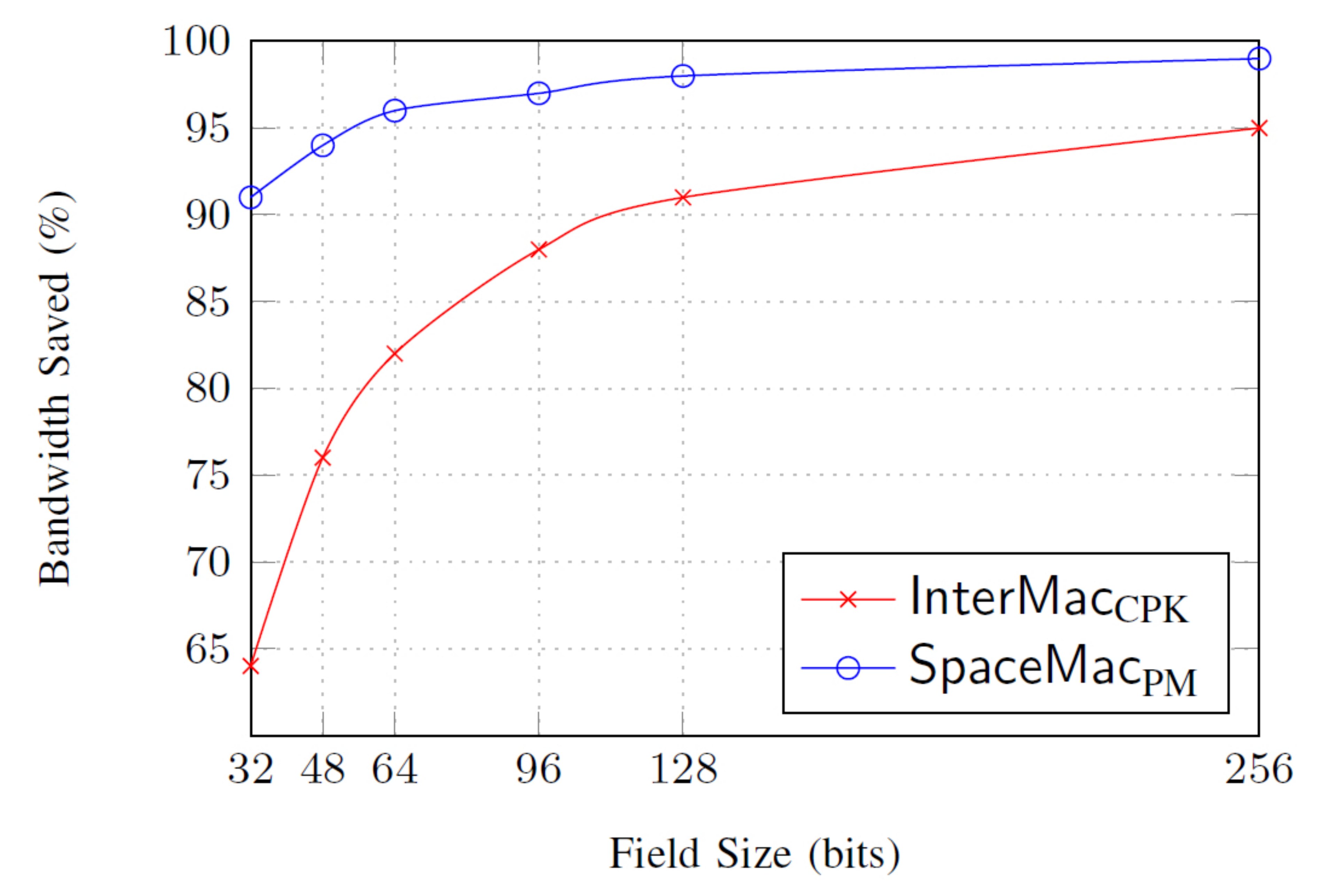}
%
%
%
\caption{Percentage of bandwidth saved with PIP as a function of field size.}
\label{fig:bandwidth}
\end{figure}

The off-line bandwidth overhead of $\IMac_\text{CPK}$ and $\SMac_\text{PM}$ come from the packets exchanged during the execution of the CPK and PM protocols. The off-line bandwidth overhead of  $\IMac_\text{CPK}$ includes the overhead of the encryptions of the randomly chosen vectors sent by the controller, the encryptions of the inner products sent back by the sources, and the padding sent by the controller, which is $ s (s-1) ( n e |q| + g e |q| )  +   s g (s-1) |q|$, where $e$ is the expansion factor of the encryption scheme and equals $\frac{N}{|q|}$ ($N$ is the size of the modulo of the encryption in bits). The off-line bandwidth overhead of $\SMac_\text{PM}$ includes the overhead of the encryption of the randomly chosen vector and the encryptions of the inner products, which is $ s ( n e |q| + g e |q| ) $. To be concrete, for $N=256$, $n=1024$, $s=5$, and $g=100$, the off-line bandwidth overhead per source packet of $\IMac_\text{CPK}$ and $\SMac_\text{PM}$ range from 36\% to 1\% as the field size increases from 32 to 256 bits. Fig. \ref{fig:bandwidth} shows the percentage of bandwidth saved when using PIP for the commitment as opposed to the sources sending all source packets to the controller. As shown in the Fig. \ref{fig:bandwidth}, the percentage of bandwidth saved increases as the field size increases. When $|q| \geq 128$, the percentage of saving is larger than 90\% for both  $\IMac_\text{CPK}$ and $\SMac_\text{PM}$. The saving could be as much as 99\% for $\SMac_\text{PM}$ when $|q|=N=256$.

The online overhead comes from the tags accompanied with each packet. To provide end-to-end detection, using a single tag suffices. In this case, the overhead of both $\IMac_\text{CPK}$ and $\SMac_\text{PM}$ is $\frac{|q|}{n |q|} (0.1\%$ for $n=1024$). To provide in-network detection for a directed acyclic network, let one of our MAC schemes be used with the delayed key disclosure technique in RIPPLE \cite{Li2010}\footnote{When using $\IMac_\text{CPK}$, the delayed MAC keys must be verified differently, \ie, using public key verification instead of one-way key chain.}. Let $S_0$ be a virtual node which has an edge pointing toward every source node. Define a level of a node as the length of the longest path from $S_0$ to the node. Let $L$ be the maximum among the levels of the nodes. Each packet carries $L$ MAC tags initially; then one or more tags are peeled off at every node the packet goes through. The average online overhead per packet is $\frac{L |q|}{2 n |q|}\,\%$.  

\begin{figure}[t]
\centering
\includegraphics[width=8.5cm]{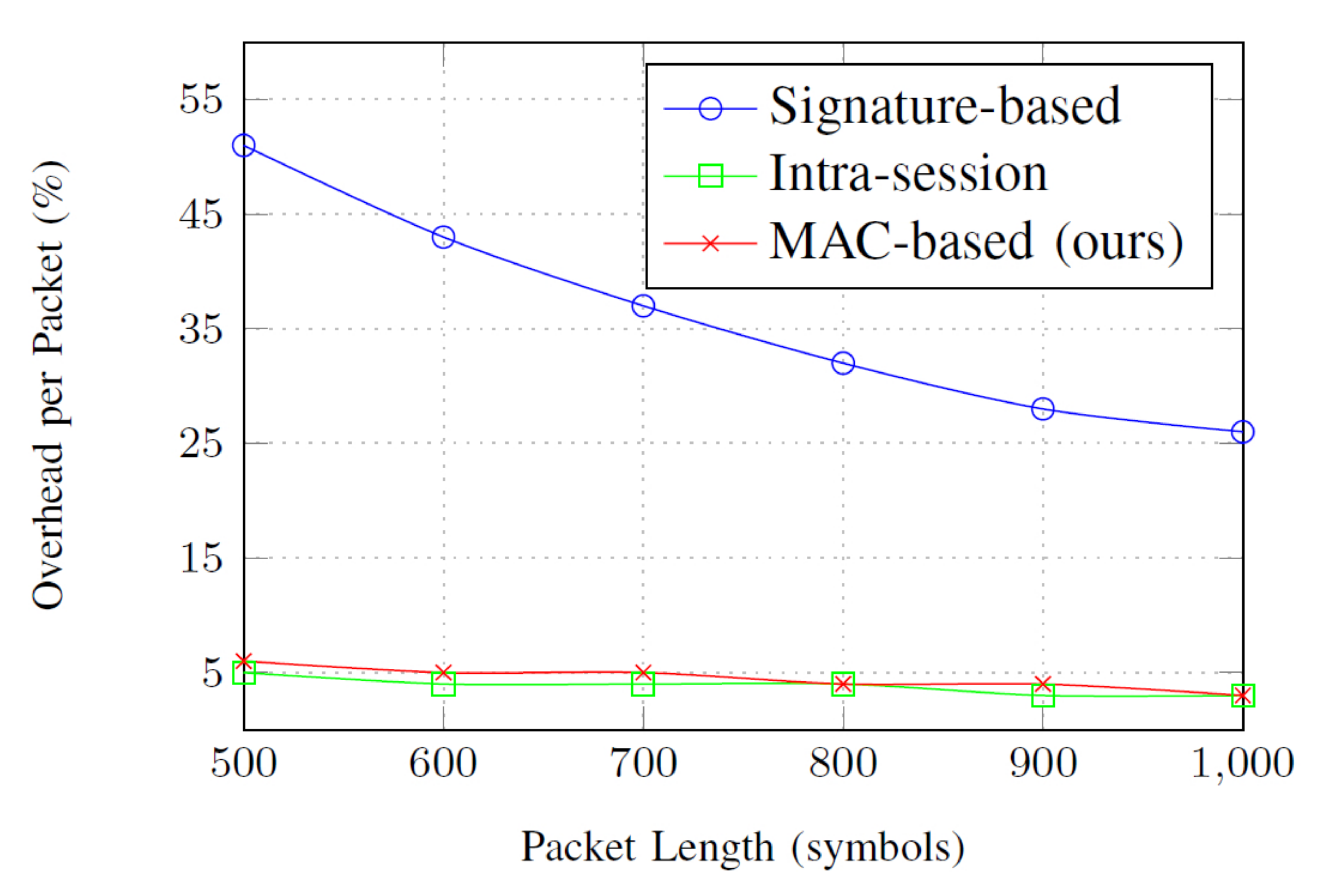}
%
%
%
%
\caption{Per-packet bandwidth overhead of the homomorphic signature scheme \cite{Agrawal2010}, the hybrid scheme in \cite{Zhang2011} ($c=1,\delta=0.1,\epsilon=0.01$), and our $\IMac_\text{CPK}$/$\SMac_\text{PM}$ when used with RIPPLE \cite{Li2010}.}
\label{fig:bandwidthOverhead}
\end{figure}

In comparison, on average, the online overhead per packet of \cite{Agrawal2010} is $\frac{s (g |q| + |\sigma|) }{2 n |q|} \,\%$, where $|\sigma|$ is the size of a regular public key signature. We stress that this overhead depends on the number of source packets whereas ours does not. To be concrete, if we set $L=16$ (as in \cite{Li2010}), $|\sigma| = 320$ ($\mathsf{DSA}$), $|q|=128$, $s=5$, $g=100$, then the overhead per packet of \cite{Agrawal2010} is $32$ times larger than ours ($\frac{s (g |q| + |\sigma|) }{L |q|}\simeq32$).  Fig. \ref{fig:bandwidthOverhead} plots the average online overhead per packet of \cite{Agrawal2010}, a state-of-the-art intra-session detection scheme \cite{Zhang2011}, and our MAC-based scheme as a function of packet length. The range of the packet length is chosen according to \cite{Zhang2011} for ease of comparison. This plot shows that not only is our overhead significantly smaller than that of \cite{Agrawal2010}, but it is also small, as small as 3\%. Our overhead is comparable to that of \cite{Zhang2011}.

\subsection{Computation Overhead}
\label{subsec:computationOverhead}

We focus on the online overhead incurred by the operation performed at each node per packet and neglect the other overhead, \eg, computing the hashes and MAC keys, as these are negligible in the number of packets in the network. Similar to \cite{Zhang2011}, we calculate the computation overhead by approximating various operations by the number of finite field multiplications. To calculate the computation time, for ease of comparison, we adopt the benchmark obtained in \cite{Zhang2011} on a 2.0 GHz Intel Core 2 CPU, where approximately $2.5 \times 10^5$ multiplications can be performed per second for $|q|$=128.

\subsubsection{Hash-Based Detection} For each packet, the worst case scenario is that the node needs to perform a homomorphic hash check, \ie, performing the \test~algorithm of \HDL. This algorithm entails $n+m$ modular exponentiations (recall $m=sg$). In comparison, in the worst case, the scheme in \cite{Agrawal2010} requires $n+m$ exponentiations plus $s$ public-key signature verifications. In the best scenario, where the received packet is decodable, our scheme just requires a traditional hash check. 

\subsubsection{MAC-Based Detection} Let one of our MAC schemes be used with RIPPLE as described in Section \ref{subsec:bandwidthOverhead}. For each packet, the overhead includes one \cbn~(to generate the tag of the packet) and one \vrf~(to verify the integrity of the packet). Let $w$ be the average number of packets combined by each node. Then, on average, the \cbn~algorithm entails $w(\frac{L-1}{2})$ multiplications; meanwhile, the \vrf~algorithm entails $n+m+\frac{L-1}{2}$ multiplications. The total average overhead is  $w(\frac{L-1}{2}) + (n+m+\frac{L-1}{2})$ multiplications.

In comparison, the average overhead of \cite{Agrawal2010} is $n + \frac{sg}{2}$ exponentiations plus $\frac{s}{2}$ public-key verification. For simplicity, approximate the cost of one public-key verification (DSA) by two modular exponentiations.  Utilizing the ``square and multiple'' method for calculating exponentiation over a finite field $\eff{}{q}$, each exponentiation over $\eff{}{q}$ takes approximately $\frac{3}{2} |q|$ multiplications on average \cite{Zhang2011}. The total average overhead is $\frac{3}{2} |q| (n + \frac{sg}{2} + s)$ field multiplications.

\begin{figure}[t]
\centering
\includegraphics[width=8.5cm]{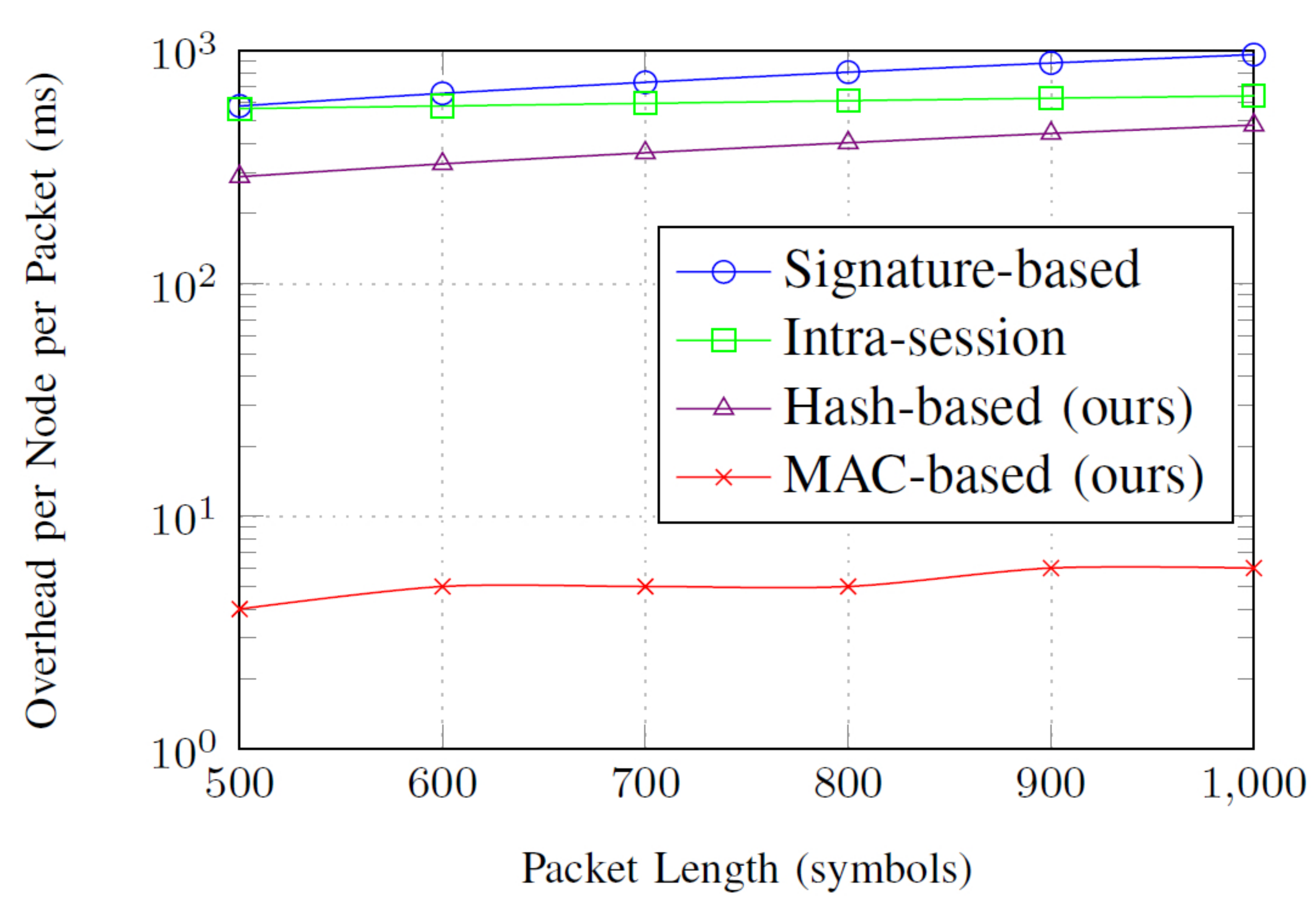}
%
%
%
%
%
\caption{Per-packet per-node computation overhead of the signature scheme \cite{Agrawal2010}, the hybrid scheme in \cite{Zhang2011} ($c=1,\delta=0.1,\epsilon=0.01$), our hash-based scheme, and $\IMac_\text{CPK}$/$\SMac_\text{PM}$ when used with RIPPLE \cite{Li2010}.}
\label{fig:computationOverhead}
\end{figure}

For concreteness, let $L=16$, $w=4$, $n=1024$, $s=5$, $g=100$, and $|q|=128$. We approximate a traditional hash check by 80 field multiplications (1 per iteration of $\SHA$) and let the decodable probability be 50\%. Fig. \ref{fig:computationOverhead} plots the average online computation overhead per packet per node of the signature-based scheme in \cite{Agrawal2010}, the intra-session detection scheme in \cite{Zhang2011}, and our hash-based and MAC-based schemes. This plot shows that the overhead of our hash-based scheme is half of that of \cite{Agrawal2010}. The computation efficiency would increase with the decodable probability. The plot also demonstrates that the overhead of our MAC-based scheme is small, ranging from 4 to 6 ms, and is two orders of magnitude less than the that of \cite{Agrawal2010} and \cite{Zhang2011}.

\section{Conclusion}
\label{sec:conclusion}
In this work, we introduce three efficient schemes to detect pollution attacks in inter-session network coding. The central idea of our schemes is the use of commitment of source packets. Our first scheme is a novel combination of homomorphic and traditional hash functions. The other two schemes are novel MAC schemes for inter-session network coding: $\IMac_\text{CPK}$ and $\SMac_\text{PM}$. To the best of our knowledge, $\IMac_\text{CPK}$ is the first multi-source homomorphic MAC scheme that support multiple keys. Except when using one-hop decoding, \eg, COPE, we recommend using detection schemes built on our MAC schemes as they have significantly lower computation overhead. Finally, we recommend using $\IMac_\text{CPK}$ over $\SMac_\text{PM}$ when there may be malicious receivers.


\balance


\end{document}